\documentclass[twocolumn,numberedappendix, twocolappendix, apj]{openjournal}
\usepackage{newtxtext,newtxmath}
\usepackage{amsmath}
\usepackage{amssymb}
\usepackage{graphicx}
\usepackage{dcolumn}
\usepackage{bm}
\usepackage[dvipsnames]{xcolor}
\usepackage{xspace}
\usepackage[colorlinks=true, allcolors=blue]{hyperref}
\usepackage{ulem}
\usepackage{lipsum}

\newcommand{\bq}{\boldsymbol q}
\newcommand{\mo}{\mathcal{O}}
\newcommand{\bx}{\boldsymbol x}
\newcommand{\bk}{\textbf{k}}

\newcommand{\bPsi}{\boldsymbol{\Psi}}

\newcommand{\ihmpc}{\,h{\rm Mpc}^{-1}}

\def\mpcoh{\,h^{-1}{\rm Mpc}} 
\def\gpcoh{\,h^{-1}{\rm Gpc}} 

\def\msunoh{\,h^{-1}{\rm M}_\odot}
\newcommand{\zenbu}{\texttt{ZeNBu}}

\begin{document}

\title{Mitigating the noise of DESI mocks using analytic control variates}

\author{B.~Hadzhiyska,$^{1,2,*}$}
\author{M.~White,$^{3,2}$}
\author{X.~Chen,$^{4}$}
\author{L.~H.~Garrison,$^{5,6}$}
\author{J.~DeRose,$^{1}$}
\author{N.~Padmanabhan,$^{4}$}
\author{C.~Garcia-Quintero,$^{7}$}
\author{J.~Mena-Fern\'andez,$^{8}$}
\author{S.~Chen,$^{9}$}
\author{H.~Seo,$^{10}$}
\author{P.~McDonald,$^{1}$}
\author{J.~Aguilar,$^{1}$}
\author{S.~Ahlen,$^{11}$}
\author{D.~Brooks,$^{12}$}
\author{T.~Claybaugh,$^{1}$}
\author{A.~de la Macorra,$^{13}$}
\author{P.~Doel,$^{12}$}
\author{A.~Font-Ribera,$^{14}$}
\author{J.~E.~Forero-Romero,$^{15,16}$}
\author{S.~Gontcho A Gontcho,$^{1}$}
\author{K.~Honscheid,$^{17,18,19}$}
\author{A.~Kremin,$^{1}$}
\author{M.~Landriau,$^{1}$}
\author{M.~Manera,$^{20,14}$}
\author{R.~Miquel,$^{21,14}$}
\author{J.~Nie,$^{22}$}
\author{N.~Palanque-Delabrouille,$^{23,1}$}
\author{M.~Rezaie,$^{24}$}
\author{G.~Rossi,$^{25}$}
\author{E.~Sanchez,$^{8}$}
\author{M.~Schubnell,$^{26,27}$}
\author{G.~Tarl\'{e},$^{27}$}
\author{M.~Vargas-Maga\~na,$^{13}$}
\author{Z.~Zhou$^{22}$}
\email{$^*$boryanah@berkeley.edu}
\affiliation{$^{1}$ Lawrence Berkeley National Laboratory, 1 Cyclotron Road, Berkeley, CA 94720, USA}
\affiliation{$^{2}$ University of California, Berkeley, 110 Sproul Hall \#5800 Berkeley, CA 94720, USA}
\affiliation{$^{3}$ Department of Physics, University of California, Berkeley, 366 LeConte Hall MC 7300, Berkeley, CA 94720-7300, USA}
\affiliation{$^{4}$ Physics Department, Yale University, P.O. Box 208120, New Haven, CT 06511, USA}
\affiliation{$^{5}$ Center for Computational Astrophysics, Flatiron Institute, 162 5\textsuperscript{th} Avenue, New York, NY 10010, USA}
\affiliation{$^{6}$ Scientific Computing Core, Flatiron Institute, 162 5\textsuperscript{th} Avenue, New York, NY 10010, USA}
\affiliation{$^{7}$ Department of Physics, The University of Texas at Dallas, Richardson, TX 75080, USA}
\affiliation{$^{8}$ CIEMAT, Avenida Complutense 40, E-28040 Madrid, Spain}
\affiliation{$^{9}$ Institute for Advanced Study, 1 Einstein Drive, Princeton, NJ 08540, USA}
\affiliation{$^{10}$ Department of Physics \& Astronomy, Ohio University, Athens, OH 45701, USA}
\affiliation{$^{11}$ Physics Dept., Boston University, 590 Commonwealth Avenue, Boston, MA 02215, USA}
\affiliation{$^{12}$ Department of Physics \& Astronomy, University College London, Gower Street, London, WC1E 6BT, UK}
\affiliation{$^{13}$ Instituto de F\'{\i}sica, Universidad Nacional Aut\'{o}noma de M\'{e}xico,  Cd. de M\'{e}xico  C.P. 04510,  M\'{e}xico}
\affiliation{$^{14}$ Institut de F\'{i}sica d’Altes Energies (IFAE), The Barcelona Institute of Science and Technology, Campus UAB, 08193 Bellaterra Barcelona, S\
pain}
\affiliation{$^{15}$ Departamento de F\'isica, Universidad de los Andes, Cra. 1 No. 18A-10, Edificio Ip, CP 111711, Bogot\'a, Colombia}
\affiliation{$^{16}$ Observatorio Astron\'omico, Universidad de los Andes, Cra. 1 No. 18A-10, Edificio H, CP 111711 Bogot\'a, Colombia}
\affiliation{$^{17}$ Center for Cosmology and AstroParticle Physics, The Ohio State University, 191 West Woodruff Avenue, Columbus, OH 43210, USA}
\affiliation{$^{18}$ Department of Physics, The Ohio State University, 191 West Woodruff Avenue, Columbus, OH 43210, USA}
\affiliation{$^{19}$ The Ohio State University, Columbus, 43210 OH, USA}
\affiliation{$^{20}$ Departament de F\'{i}sica, Serra H\'{u}nter, Universitat Aut\`{o}noma de Barcelona, 08193 Bellaterra (Barcelona), Spain}
\affiliation{$^{21}$ Instituci\'{o} Catalana de Recerca i Estudis Avan\c{c}ats, Passeig de Llu\'{\i}s Companys, 23, 08010 Barcelona, Spain}
\affiliation{$^{22}$ National Astronomical Observatories, Chinese Academy of Sciences, A20 Datun Rd., Chaoyang District, Beijing, 100012, P.R. China}
\affiliation{$^{23}$ IRFU, CEA, Universit\'{e} Paris-Saclay, F-91191 Gif-sur-Yvette, France}
\affiliation{$^{24}$ Department of Physics, Kansas State University, 116 Cardwell Hall, Manhattan, KS 66506, USA}
\affiliation{$^{25}$ Department of Physics and Astronomy, Sejong University, Seoul, 143-747, Korea}
\affiliation{$^{26}$ Department of Physics, University of Michigan, Ann Arbor, MI 48109, USA}
\affiliation{$^{27}$ University of Michigan, Ann Arbor, MI 48109, USA}
\date{\today}

\begin{abstract}
In order to address fundamental questions related to the expansion history of the Universe and its primordial nature with the next generation of galaxy experiments, we need to model reliably large-scale structure observables such as the correlation function and the power spectrum. Cosmological $N$-body simulations provide a reference through which we can test our models, but their output suffers from sample variance on large scales. Fortunately, this is the regime where accurate analytic approximations exist. To reduce the variance, which is key to making optimal use of these simulations, we can leverage the accuracy and precision of such analytic descriptions using Control Variates (CV). 
The power of control variates stems from utilizing inexpensive but highly correlated surrogates of the statistics one wishes to measure. The stronger the correlation between the surrogate and the statistic of interest, the larger the variance reduction delivered by the method.
We apply two control variate formulations to mock catalogs generated in anticipation of upcoming data from the Dark Energy Spectroscopic Instrument (DESI) to test the robustness of its analysis pipeline. Our CV-reduced measurements, of the power spectrum and correlation function, both pre- and post-reconstruction, offer a factor of 5-10 improvement in the measurement error compared with the raw measurements from the DESI mock catalogs. We explore the relevant properties of the galaxy samples that dictate this reduction and comment on the improvements we find on some of the derived quantities relevant to Baryon Acoustic Oscillation (BAO) analysis. We also provide an optimized package for computing the power spectra and other two-point statistics of an arbitrary galaxy catalog as well as a pipeline for obtaining CV-reduced measurements on any of the \textsc{AbacusSummit} cubic box outputs. We make our scripts, notebooks, and benchmark tests against existing software publicly available and report a speed improvement of a factor of $\sim$10 for a grid size of $N_{\rm mesh} = 256^3$ compared with \texttt{nbodykit}.
\end{abstract}

\maketitle

\section{Introduction}

The quest to constrain the initial perturbations, the growth of cosmic structure, and the cosmic expansion history has spawned a number of large-scale structure experiments. These include the Dark Energy Spectroscopic Instrument \citep[DESI;][]{2016arXiv161100036D,2019BAAS...51g..57L}, the Euclid space telescope \citep{2011arXiv1110.3193L,2013LRR....16....6A,2022A&A...662A.112E}, the Rubin Observatory Legacy Survey of Space and Time \citep[LSST;][]{2012arXiv1211.0310L,2019ApJ...873..111I}, the Nancy Grace Roman Space Telescope \citep{2015arXiv150303757S}, the Prime Focus Spectrograph \citep[PFS;][]{2016SPIE.9908E..1MT}, the Spectro-Photometer for the History of the Universe, Epoch of Reionization, and Ices Explorer \citep[SPHEREx;][]{2014arXiv1412.4872D,2018arXiv180505489D}, and WEAVE-QSO \citep{2016sf2a.conf..259P}. The exquisitely detailed maps of the Universe these data will allow us to pin down possible deviations from the standard paradigm and reveal the nature of its most elusive ingredients. To unlock the true potential of these new generation of surveys and meet their science goals, it is crucial to match our experimental efforts with theoretical ones.

Purely analytical models of the large-scale structure statistics such as standard perturbation theory \citep[SPT;][]{1986ApJ...311....6G,1994ApJ...431..495J}, Lagrangian perturbation theory \citep[LPT;][]{Buchert87,Bouchet95,Matsubara08,Carlson13}, renormalised perturbation theory \citep{2006PhRvD..73f3519C} and effective field theory \citep[EFT;][]{2012JHEP...09..082C,2015JCAP...09..014V,2016arXiv161009321P}, have made great strides towards accurately describing the small-scale distribution of structure \citep[see e.g.,][]{Ivanov22}. However, it is still the case that their reference models are based on computationally intensive cosmological $N$-body simulations, which provide a numerical window into the complex non-linear regime of structure growth. Examples of these sophisticated simulation suites involve the MICE Grand Challenge \citep{2015MNRAS.448.2987F,2015MNRAS.453.1513C,2015MNRAS.447.1319F,2015MNRAS.447..646C}, the Abacus Cosmos suite \citep{2018ApJS..236...43G}, the Outer Rim Simulation \citep{2019ApJS..245...16H}, the Aemulus project I \citep{2019ApJ...875...69D}, the BACCO simulation project \citep{2021MNRAS.507.5869A}, the \textsc{AbacusSummit} $N$-body simulation suite \citep{2021MNRAS.508.4017M}, the Cardinal mock catalogs\citep{2023arXiv230312104T}, the Aemulus $\nu$ project \citep{2023arXiv230309762D}, and the Euclid flagship simulation \citep{2019MNRAS.484.5509E}. While analytical methods calculate expectation values of large-scale structure statistics (i.e., with no noise), a simulation generates a single realisation, so its output suffers from sample variance. Reducing the sample variance below the observational error would traditionally require running a large number of these simulations, such that their total volume exceeds that of the experiment of interest.

An alternative to running an ensemble of $N$-body simulations to reduce sample variance is offered via the control variates technique. The power of control variates stems from utilizing inexpensive but highly correlated surrogates of the statistics one wishes to measure. The stronger the correlation between the surrogate and the statistic of interest, the larger the variance reduction delivered by the method. Thus, the optimal use of control variates is predicated on a comprehensive understanding of the statistics of the surrogate; namely, its mean, variance and co-variance. 

The method of control variates has recently received significant attention in the context of cosmology. The first applications have involved the use of approximate simulations, which due to their relatively cheap numerics, yield considerable gains in reducing the noise of the desired statistic \citep{2016MNRAS.463.2273F,2021MNRAS.503.1897C,2022MNRAS.509.2220C,2022MNRAS.514.3308D}. A benefit of this method is that the simulation initial conditions do not need to be altered, e.g.,\ by generating inverse-phase modes or introducing non-Gaussianity. However, one still needs outputs from hundreds of these approximate simulations in order to estimate the mean, variance and co-variance of the surrogate. In addition, an important issue is that if the mean is biased, as would be the case if not a sufficient number of simulations have been run, then the CV estimate would also be biased. The error bar shrinks by $\sqrt{N_{\rm sim}}$, where $N_{\rm sim}$ is the number of simulations. Thus, even if the 100 approximate simulations are $100\times$ faster than the nominal one, one only gains an improvement of a factor of 10 for double the total compute time. To get $100\times$ reduction, we would need to run $10^4$ simulations for a $10^2\times$ the cost of the original sim. Achieving ultra-high precision on the measurements using this method thus requires substantial computational efforts.

An alternative surrogate that does not suffer from the issue of computational expense is supplied by analytic descriptions of summary statistics, which on large scales are highly accurate and precise. Recently, \citet{Kokron22b} investigated the Zeldovich approximation \citep[ZA;][i.e., first-order LPT]{Zeldovich70} as a surrogate and dubbed it Zel’dovich control variates (ZCV). A benefit of ZCV is that the mean prediction is known analytically and that the ZA matter density fields for the same initial conditions seed are strongly correlated with the late-time $N$-body density fields \citep{Doroshkevich80,Coles93,Pauls95}. Follow-up work by \citet{DeRose23}, further extends the ZA and ZCV method to analytically predict the means of summary statistics in redshift space. Both papers find a reduction in the variance of 10$^2$ to 10$^6$ for a number of different tracers at $k < 0.2 \ihmpc$. The ZCV technique was also used in the construction of the emulator for the Aemulus $\nu$ project \citep{2023arXiv230309762D}.

The focus on the present study is in applying the CV technique to realistic galaxy mock catalogs prepared for the early data analysis of the DESI survey. The main advantage of mitigating the error on the measured summary statistics to early DESI science is generating low-noise data vectors for calibrating systematic errors and testing the robustness of the analysis pipeline. We release a robust and easy-to-use package for applying ZCV reduction to mocks generated with the \textsc{AbacusSummit} suite. We also test the performance of our code against the widely used \texttt{nbodykit} package \citep{2018AJ....156..160H}. Novel in this work is the development of the ZCV method for the configuration-space two-point correlation function as well as the development of the adjacent Linear control variates (LCV) method, used to reduce the noise of two-point statistics computed with the `reconstructed' density fields.

The reconstruction of the density field offers a means of improving the distance-redshift relation via the analysis of the baryon acoustic oscillations \citep[BAO;][]{Dodelson20}. The BAO peak is a feature in the 2-point function whose physical size is known. Comparisons of the observed size of the peak with its physical size yield the angular diameter distance and Hubble parameter as a function of redshift. Due to nonlinear evolution, the oscillations on small scales are dampened, which broadens the peak and reduces the accuracy with which its size can be measured. Much of the peak broadening, however, is sourced by very long wavelength fluctuations, occurring on the largest scales \citep{ESW07}. For this reason, all recent BAO surveys (including DESI) apply density-field reconstruction \citep{ESSS07,Padmanabhan12} to regain some of the information lost due to the large-scale displacements. Hence, reducing the noise on the reconstructed measurement of the correlation function and the power spectrum is of importance to large-scale structure spectroscopic experiments.

This paper is structured as follows. In Section~\ref{sec:back}, we review the control variates technique and its application on cosmology via the ZA, extending the ZCV method to configuration space. In Section~\ref{sec:lcv}, we then provide the theoretical model for the LCV and detail the steps for diminishing the noise on the reconstructed correlations. In Section~\ref{sec:app}, we illustrate the effect of ZCV and LCV on realistic mocks used in the analysis of DESI. Additionally, we comment on the improvement of BAO parameter constraints and compare it with the predicted improvement from the Fisher forecast. Finally, we summarize our main results in Section~\ref{sec:conc}.

\section{Background}
\label{sec:back}

In this Section, we provide a short overview of the control variates method and its application to cosmology via the Zeldovich approximation. We also describe the process of obtaining noise-reduced configuration-space measurements of the two-point correlation function.

\subsection{Basic concept}
\label{sec:basic}

\begin{figure}
    \centering
    \includegraphics[width=0.48\textwidth]{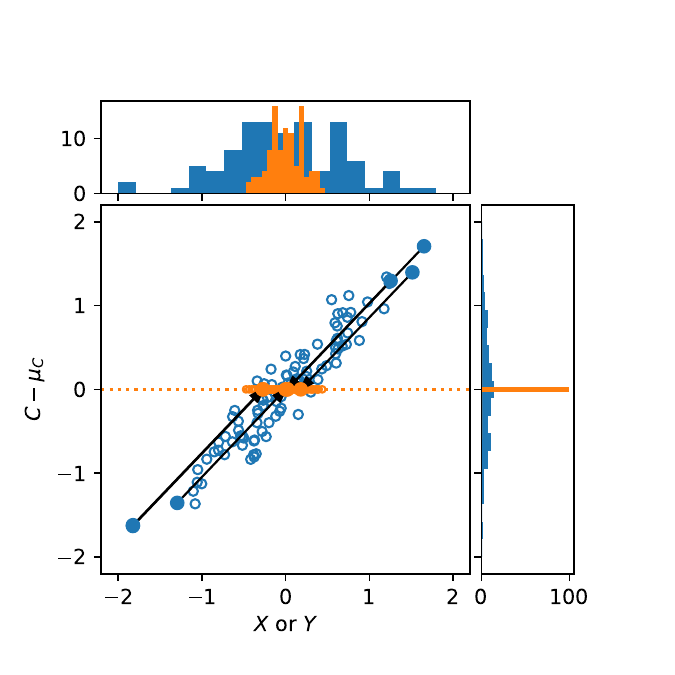} 
    \caption{Illustration of control variates. The variables $X$ and $C$ are correlated and the mean of $C$ is known, $\mu_C$. The unfilled blue circles represent random samples drawn from $X$ and $C$. Shifting the blue (filled) circles $(X,C)$ along the degeneracy direction (arrows) to the orange points $(Y,\mu_C)$ gives smaller scatter in $Y$ than originally in $X$, thus enabling a more accurate measure of the mean (here 0) from a finite number of simulations.} 
    \label{fig:cv}
\end{figure}

The control variates technique is a powerful tool in statistics for reducing the variance of a random variable, $X$, given another correlated random variable, $C$, with known mean, $\mu_c$ \citep{mcbook}. We can then write an estimator for a new variable, $Y$, as follows:
\begin{align}
    \label{eq:cv}
    Y = X - \beta (C - \mu_c),
\end{align}
\noindent 
for some arbitrary coefficient, $\beta$. Provided that $\langle C - \mu_c \rangle=0$, this is an unbiased estimator regardless of what $\beta$ is set to, as long as $\beta$ and $C$ are uncorrelated. If we wish to minimize the variance of $Y$, it can be shown \citep[see e.g.,][]{2022MNRAS.509.2220C} that the optimal choice for $\beta$ is
\begin{align}
\label{eq:beta}
\beta^{\star}=\frac{\textrm{Cov}[X,C]}{\textrm{Var}[C]},
\end{align}
\noindent 
where $\textrm{Cov}$ and $\textrm{Var}$ denote the statistical covariance and variance between the variables.
In this case, the variance of $Y$ is given by
\begin{align}
    \textrm{Var}[Y] &= \textrm{Var}[X]\left(1 - \frac{\textrm{Cov}^2[X,C]}{\textrm{Var}[C]\textrm{Var}[X]} + \beta^{2}\textrm{Var}[\mu_c]\right),
\end{align}
\noindent
where $\textrm{Var}[\mu_c]$ is the uncertainty on the mean of $C$. We note that here we implicitly assume that $\mu_c$, $X$ and $\beta$ are not correlated with each other. Often, $\mu_c$ is estimated via Monte Carlo realizations of $C$, which then can incur a large computational cost on the control variate reduction. On the other hand, if $\mu_c$ is known analytically, then we can circumvent the computational expense. Moreover, for analytic quantities the variance vanishes, and the above expression simplifies to:
\begin{align*}
    \textrm{Var}[Y] = \textrm{Var}[X](1 - \rho_{xc}^2), 
\end{align*}
where $\rho_{xc}$ denotes the cross-correlation coefficient between $X$ and $C$:
\begin{align*}
    \rho_{xc} \equiv \frac{\textrm{Cov}[X,C]}{\sqrt{\textrm{Var}[C]\textrm{Var}[X]}} .
\end{align*}
Thus, $\sqrt{(1 - \rho_{xc}^2)}$ gives us the reduction in noise on the quantity of interest after applying CV. The basic principle of control variates is demonstrated in Fig.~\ref{fig:cv}. Given a random variable, $X$, correlated with $C$, for which we draw a finite number of samples, we can move the samples along the degeneracy direction towards the known mean of $C$, $\mu_c$, and obtain new samples, $Y$, with a better-measured mean (i.e., less noisy).

Relevant to cosmology is the case where $X$ is a measurement of a summary statistic such as the power spectrum or correlation function, calculated from an $N$-body simulation. 
\citet{Kokron22b} and \citet{DeRose23} showed that the ZA is an excellent choice for the quantity $C$, as it is highly correlated with the late-time density field and has an analytically known mean in both real and redshift space.

\subsection{Review on Zeldovich control variates}
\label{sec:review}

In the Lagrangian picture, we are interested in the time evolution of phase-space elements labeled by their initial positions, $\bq$. The displacements $\Psi(\bq,a)$ then allow us to calculate their positions, $\bx$, at some later time, $a$, via $\bx = \bq + \Psi(\bq,a)$. The density of a biased tracer $\delta_t$ can be expressed as
\begin{equation}
    1 + \delta_t(\bx,a) = \int d^3\bq \ F(\bq)\ \delta_D(\bx - \bq - \Psi(\bq,a)),
    \label{eqn:density_field}
\end{equation}
where $\delta_D$ is the Dirac delta function and
\begin{equation}
    F(\bq) = 1 + b_1 \delta_0 + b_2 \big( \delta_0(\bq)^2 - \langle \delta_0^2 \rangle \big) + b_{s^2} \big( s_0(\bq)^2 - \langle s_0^2 \rangle \big) + ...
    \label{eqn:bias_expansion}
\end{equation}
with $\delta_0$ the linear density field and $s^2 = s_{ij} s_{ij}$ the local shear field \citep{Matsubara08,Chen21}. 

The displacements, $\Psi$, can be computed either analytically or numerically, through the use of $N$-body simulations. The latter approach is perhaps the most straightforward: in an $N$-body simulation, the Lagrangian position $\bq$ is the initial grid position of the particle and, and $\Psi$ is the vector connecting that initial position with a position at some later time. However, because the largest contribution to $\Psi$ comes from large-scale bulk flows \citep{ESW07} that are very well captured by the ZA, one can simply advect the particles by the Zeldovich displacement and capture the decorrelation of large scale structure to leading order \citep{Bouchet95,Buchert89,Pauls95,Yoshisato06,Tassev14,White14,2019PhRvD.100b3543C}. In the ZA, it can be shown that the real-space displacement is given by 
\begin{equation}
    \Psi^{(1)}(\bk,a) = \frac{i \bk}{k^2}\  D(a)\ \delta_0(\bk),
    \label{eq:za_displ}
\end{equation}
where $D(a)$ is the linear-theory growth factor and $\delta_0(\bk)$ is the linear density, while the redshift-space displacements can simply be obtained by multiplying the real-space ones by a constant matrix \citep{Matsubara08}
\begin{equation}
    \Psi^{(1)}_{s} = \mathbf{R} \Psi^{(1)}, \quad R_{ij} = \delta_{ij} + f \hat{n}_i \hat{n}_j,
    \label{eq:zspace_za_displ}
\end{equation}
where $\hat{n}$ is the line-of-sight unit vector,  $s$ denotes redshift space, and $f = d\ln D /d\ln a$ is the linear growth rate. We note that $R_{ij}$ takes the form above only in the plane-parallel approximation, but when working with wide-field surveys such as DESI, this approximation no longer holds, and one needs to take into account the geometry of the survey. We can then express the biased tracer power spectrum as
\begin{equation}
    P^{tt}(\bk) = \sum_{\mo_i,\mo_j} b_{\mo_i}b_{\mo_j} P_{ij}(\bk),
    \label{eq:biased_tracer_spectra}
\end{equation}
where we have defined the \textit{basis spectra} as
\begin{equation}
    P_{ij}(\bk) (2\pi)^3 \delta_D(\bk+\bk') = \Big \langle \mo_{i}(\bk)\mo_{j}(\bk')\Big \rangle.
\end{equation}
The operators $\mo_{i}(\bk)$ can either signify the redshift- or real-space advected fields for 1 (i.e., matter), $\delta_0$, $\delta^2_0$, $s^2_0$ and $\nabla^2 \delta_0$. These operators come from invoking the equivalence principle, in terms of which the leading gravitational effects are associated with the Hessian of the gravitational potential $\partial_i \partial_j \Phi$, which can be split into a scalar trace, proportional to the matter overdensity $\delta_0$, and the traceless tidal tensor $s^0_{ij}$.

We can make predictions for the ensemble mean of these statistics with the \texttt{ZeNBu} code\footnote{\href{https://github.com/sfschen/ZeNBu}{https://github.com/sfschen/ZeNBu}} to directly compute $\mu_C$. On the other hand, we can use the $N$-body simulation to measure the control variate quantity, $C$, using the following procedure. Given a realization of the initial conditions of a simulation, one can calculate the $\delta_0^2,\, s_0^2,\, \nabla^2 \delta_0$ fields from the linear density field, $\delta_0$ as well as the displacement field using Eq.~\ref{eq:za_displ} and Eq.~\ref{eq:zspace_za_displ}. At a given epoch of interest, we can then perform the advection integral, i.e. Eq.~\ref{eqn:density_field}, numerically and compute the basis spectra, i.e., the cross-correlations between the fields. For a choice of bias parameters, $b_1,\, b_2,\, b_{s^2}\,$, we can then predict the tracer power spectra by summing the terms in Eq.~\ref{eq:biased_tracer_spectra} and multiplying them by the biases. 

\begin{figure*}
    \centering
    \includegraphics[width=0.33\textwidth]{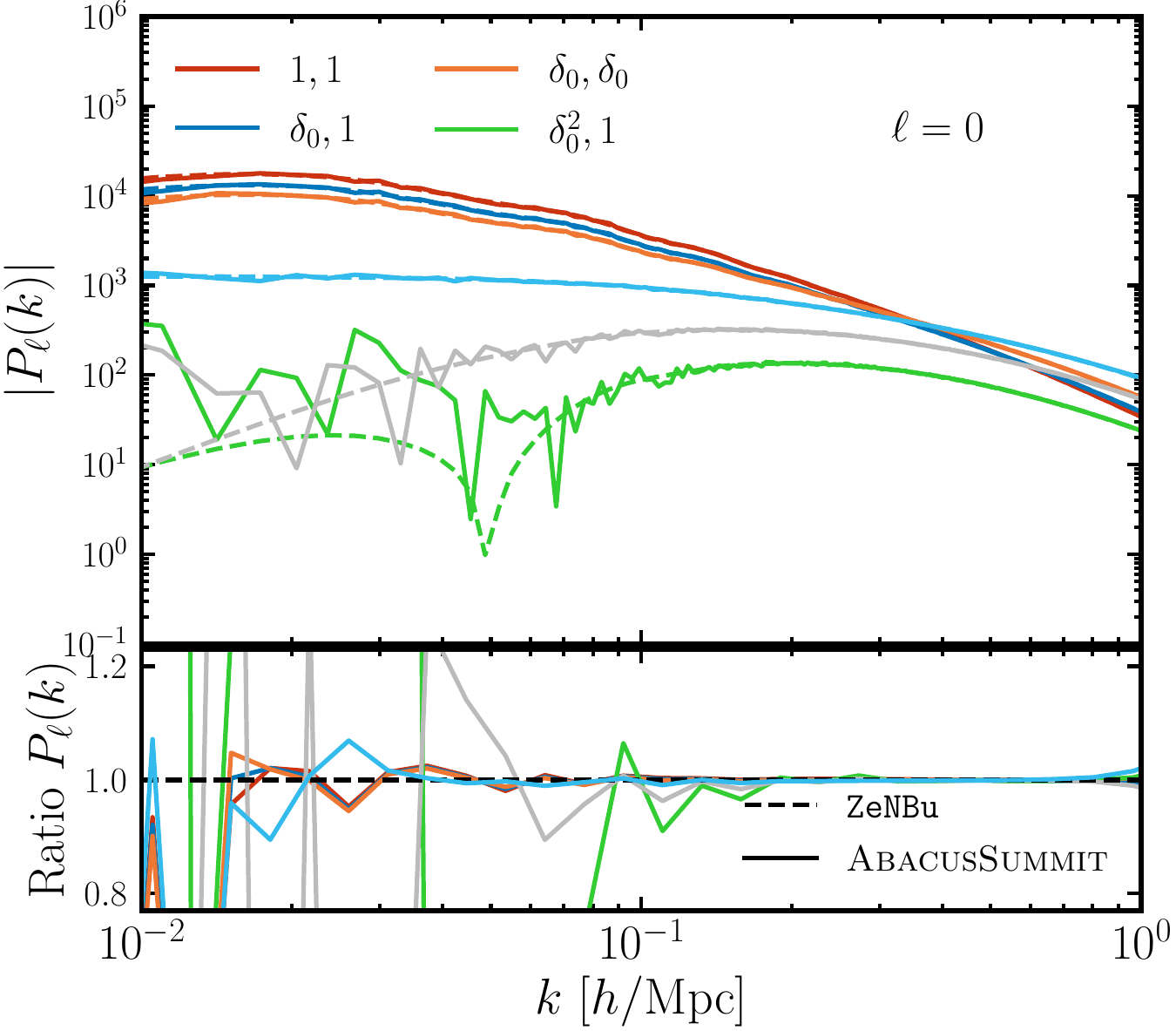}
    \includegraphics[width=0.32\textwidth]{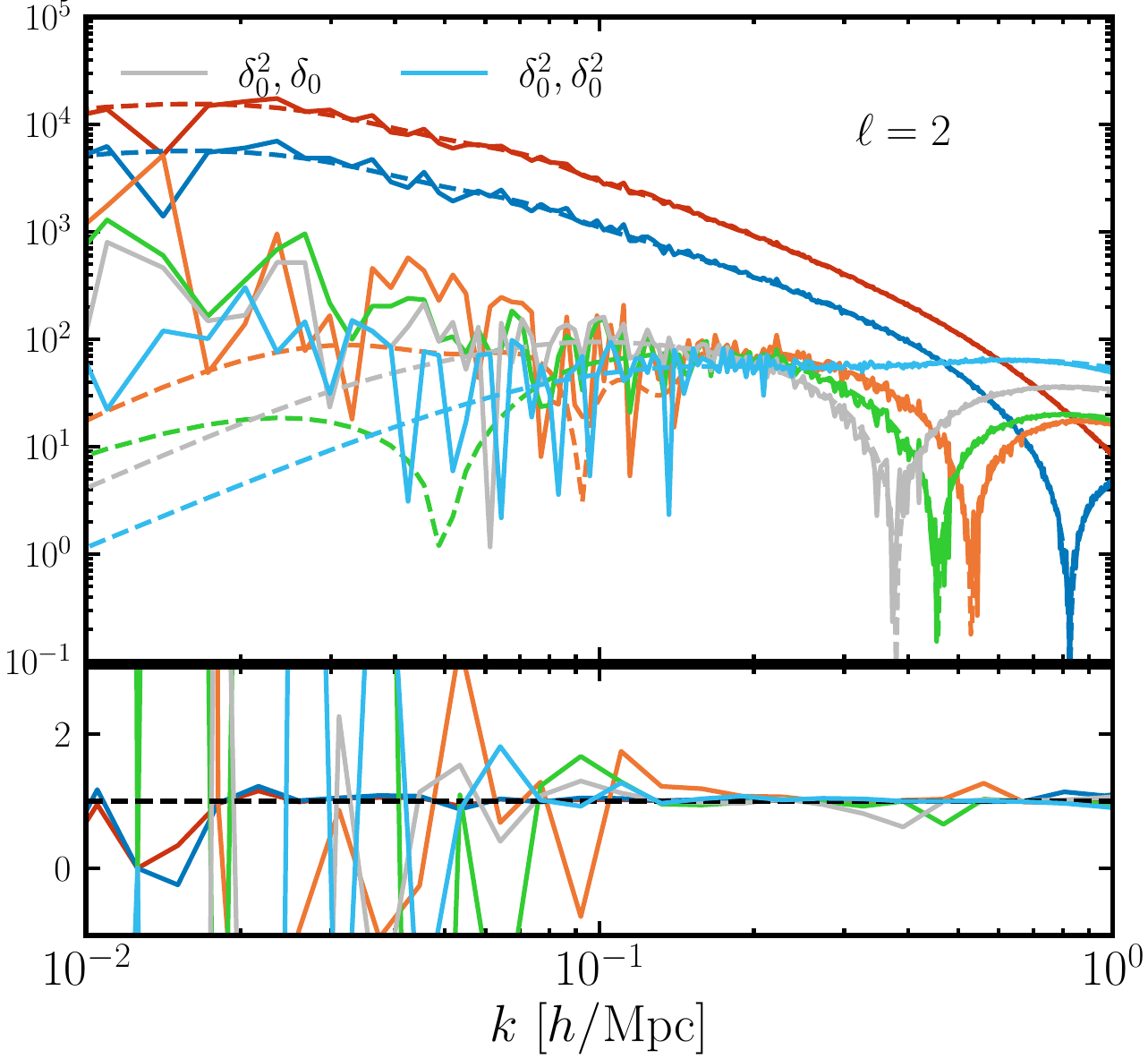}
    \includegraphics[width=0.32\textwidth]{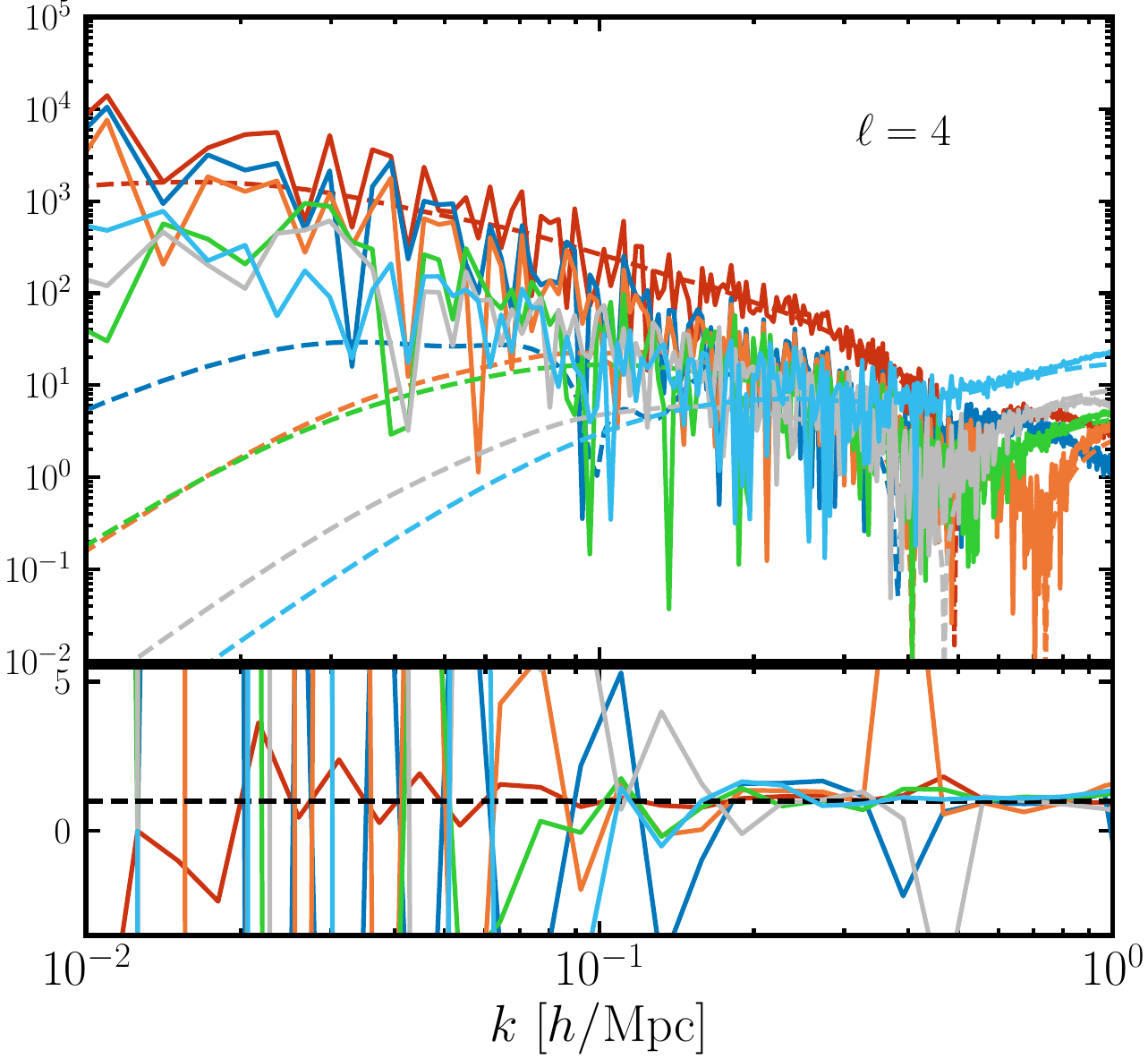}
    \caption{Basis spectra of the Zeldovich approximation in redshift space (see Section~\ref{sec:back}) calculated using \zenbu\ (dashed lines) and the \texttt{AbacusSummit\_base\_c000\_ph000} simulation (solid lines) at $z = 0.8$. On large scales, especially for the higher-order fields, the measurements become very noisy, as the signal is dominated by sample variance. Power spectra including the $\nabla^2 \delta_0$ field are omitted as they are well-approximated by $P_{X \nabla^2 \delta_0}(k) \approx k^2 P_{X \delta}$. Central to the ZCV method is ensuring that the theoretical prediction is unbiased with respect to the numerical result (for more discussion and validation, see \citealt{Kokron22b}).} 
    \label{fig:pk_ij}
\end{figure*}

\citet{Kokron22b} and \citet{DeRose23} showed that smoothing $\delta_0$ using a Gaussian kernel of scale $k_{\rm smooth}=\pi \, N_{\rm mesh}/(2L_{\rm box})$, where $N_{\rm mesh}$ is the grid size and $L_{\rm box}$ the box size, improves the agreement between the ensemble average of the grid-based realizations and the analytic prediction, which is needed to validate the ZCV method. We refer the reader to these two works for discussions on the accuracy of the analytical and numerical calculations. In Fig.~\ref{fig:pk_ij}, we illustrate the agreement between \zenbu\ and \textsc{AbacusSummit} in redshift space for a single realization, finding it to be sufficient given the numerical noise.

We use the optimized package \texttt{abacusutils}\footnote{\href{https://github.com/abacusorg/abacusutils}{https://github.com/abacusorg/abacusutils}} to compute the simulation power spectrum and correlation function, and the \textsc{AbacusSummit} suite products for the linear density field and Zeldovich displacements as well as the late time snapshot outputs. We also provide notebooks and examples for how to use the control variates code and power spectrum \href{https://github.com/abacusorg/abacusutils/tree/power_spec/docs/tutorials/analysis}{at this URL}. We discuss optimization in App.~\ref{app:python}.

\subsection{From power spectra to correlation functions}
\label{sec:power2xi}

Of most interest to this study is the redshift-space application of the Linear and Zeldovich control variates, as our main goal is to aid the high-precision analysis of redshift surveys. \citet{DeRose23} presented the ZCV formalism for mitigating the noise on the power spectrum multipoles. Here, we offer an extension to configuration space via the correlation function multipoles.

The procedure of \citet{DeRose23} for obtaining the ZCV-reduced variance tracer power spectrum, $\hat{P}^{\ast, tt}_{\ell}(k)$, in redshift-space involves the following calculation:
\begin{equation}
    \hat{P}^{\ast, tt}_{\ell}(k) = \hat{P}^{tt}_{\ell}(k) - \beta_{\ell}(k) \left(\hat{P}^{ZZ}_{\ell}(k) - P^{ZZ}_{\ell}(k)\right) ,
\end{equation}
where $\hat{P}^{tt}_{\ell}(k)$ is the measured power spectrum from the simulation, $\hat{P}^{ZZ}_{\ell}(k)$ is the measured ZA tracer power spectrum, and $P^{ZZ}_{\ell}(k)$ is the ensemble-average ZA tracer power spectrum. We note that following \citet{DeRose23}, we account for multipole mixing in the \zenbu\ prediction due to the discrete $\mu$ sampling of the measurements. The ZA quantities are computed using the advected fields and the basis spectra (see Eq.~\ref{eq:biased_tracer_spectra}). The `ensemble-average' quantities are derived analytically via \zenbu. When working in real space, we can drop the $\ell$ subscripts and work with the isotropic power spectrum, $P(k)$. As noted in Eq.~\ref{eq:cv}, the quantities $\beta$, $\mu_C$ and $X$ are assumed to be uncorrelated. In our case, this holds as $\mu_C \equiv P_\ell^{ZZ} (k)$ is analytic, and $\beta \equiv \beta_\ell(k)$ is smoothed to remove any noise from its estimation.

We find that when performing fits to the measured power spectrum to determine the bias parameters, fitting beyond $b_1$ makes very little difference to the ZCV performance. An alternative would be to fit the bias parameters at the field level, but the returns of this more involved procedure are diminishing and the gains insignificant. We note that \citet{DeRose23} find that including $b_2$ improves the CV cross-correlation coefficient by $\sim$20\% in the case of higher-bias samples. In addition, the ZA is not as accurate on intermediate scales as higher-order LPT, and hence the fit is not as stable. 
For this reason, we model the Zeldovich predictions by including only the first-order bias:
\begin{equation}
    \hat P^{ZZ}_{\ell}(k) = P^{mm}_{\ell}(k) + 2 b_1 P^{1\delta_0}_{\ell}(k) + b_1^2 P^{\delta_0\delta_0}_{\ell}(k),
\end{equation}
which we obtain by minimizing the difference:
\begin{equation}
    \sum_{k<k_{\rm max}} \left(\hat{P}^{ZZ}_{\ell=0}(k) - \hat{P}^{tt}_{\ell=0}(k) \right)^2,
\end{equation}
where we have set $k_{\rm max}=0.15\ihmpc$ following \citet{DeRose23}. We weight each $k$-bin equally in the optimization function (as opposed to the standard $\chi^2$, which gives more weight to the higher $k$-bins that have more modes and hence smaller errors) in order to upweight the large scales, for which the ZA is more accurate.

To estimate $\beta_{\ell}(k)$, which is defined as:
\begin{equation}
    \beta_{\ell}(k) = \frac{\textrm{Cov}[\hat{P}^{tt}_{\ell}(k), \hat{P}^{ZZ}_{\ell}(k)]}{\textrm{Var}[\hat{P}^{ZZ}_{\ell}(k)]},
\end{equation}
we adopt the disconnected approximation, such that:
\begin{equation}
    \beta_{\ell}(k) = \left[ \frac{\hat{P}^{tZ}_{\ell}(k)}{\hat{P}^{ZZ}_{\ell}(k)} \right]^2,
\end{equation}
where $\hat{P}^{tZ}(k)$ is the measured cross-power spectrum between the tracer in question and our ZA control variate and $\hat{P}^{ZZ}(k)$ is the auto-power spectrum of the ZA control variate (see App. C of \citet{DeRose23} for a discussion of the accuracy of the disconnected approximation). We note that $\beta_\ell(k)$ is additionally damped with a $\tanh$ and smoothed with a Savitsky-Golay filter in order to ensure that $\beta_\ell(k)$ is uncorrelated with the measured power spectrum and that it approaches zero where the disconnected approximation breaks down (see App. C of \citealt{DeRose23} for more details).

In order to obtain the ZCV-reduced correlation function, it is natural to work with the three-dimensional power spectrum, $P(\bk)$, and transform the final quantity into configuration space. The reason we opt not to apply a Hankel transformation to the power spectrum multipoles is that this procedure is very sensitive to the smoothness of the function being transformed, and we aim to avoid artificially modifying the measurement.
Thus, the ZCV equation becomes:
\begin{equation}
    \label{eq:zcv_xi}
    \hat{P}^{\ast, tt}(\bk) = \hat{P}^{tt}(\bk) - \beta(\bk) \left(\hat{P}^{ZZ}(\bk) - P^{ZZ}(\bk)\right) .
\end{equation}
While the measured quantities can be directly estimated as $\hat{P}(\bk) = |\delta(\bk)|^2$, for $\beta(\bk)$ and $P^{ZZ}(\bk)$, we need to construct three-dimensional objects using the Legendre polynomials:
\begin{align}
    \label{eq:beta_leg}
    &\beta(k, \mu) = \sum_{\ell = 0,\,2,\,4} \beta_\ell(k) \mathcal{L}(\mu) \\
    &P^{ZZ}(k, \mu) = \sum_{\ell = 0,\,2,\,4} P^{ZZ}_{\ell}(k) \mathcal{L}(\mu)
    \label{eq:ZZ_leg}
\end{align}
and evaluate them on the three-dimensional Fourier grid, $\bk = \{k_x,\,k_y,\,k_z\}$, assuming azimuthal symmetry around the $z$ (line-of-sight) axis with $\mu = |k_z|/k$. We make these estimates via the multipoles in order to reduce the noise so that $\beta$ is smooth and uncorrelated with $\hat{P}^{tt}(\bk)$ and $\hat{P}^{ZZ}(\bk)$.

\begin{figure}
    \centering
    \includegraphics[width=0.48\textwidth]{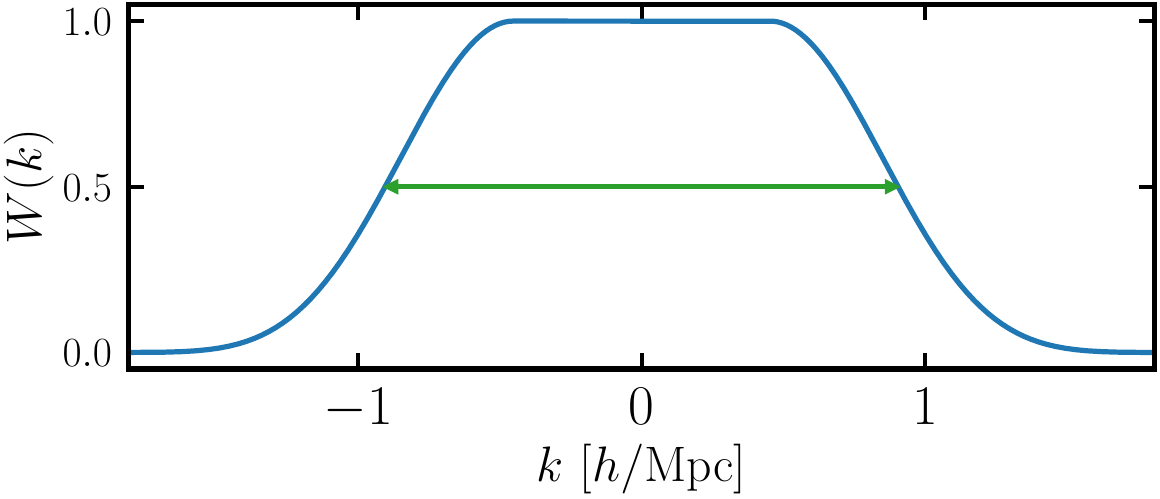}
    \includegraphics[width=0.48\textwidth]{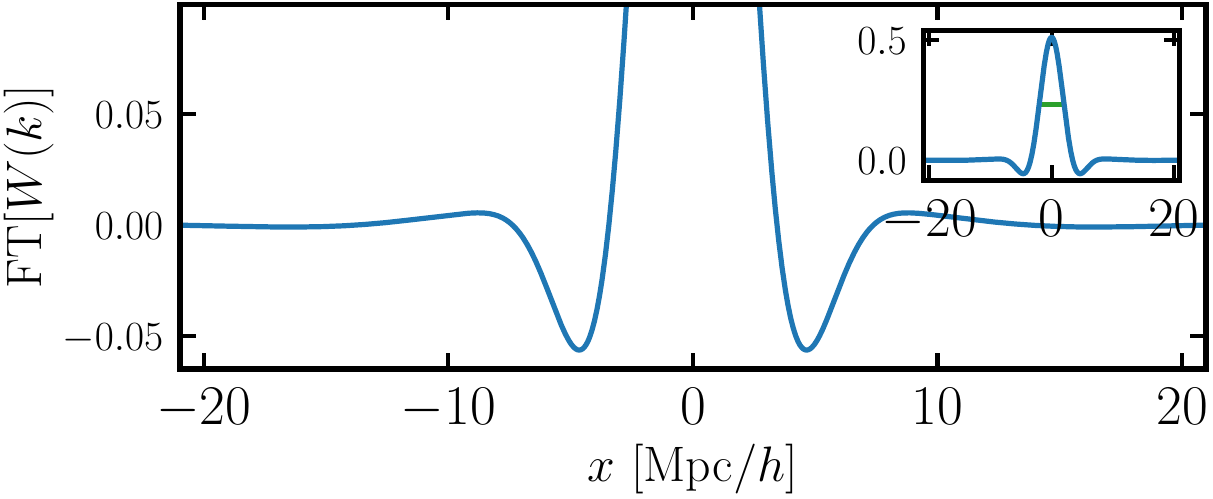} 
    \caption{Blackman-Harris apodizing (tapering) window function (defined in Eq.~\ref{eq:apod}) in Fourier (top panel) and in configuration space (bottom panel). We apply this tapering function to the three-dimensional power spectrum $P(\bk)$ before inverse-Fourier transforming it into a correlation function, $\xi(\bx)$. The full width at half maximum (FWHM) in Fourier space is roughly $0.68636\, a$, which for our settings corresponds to the Nyquist frequency, i.e., ${\rm FWHM} \approx 1.8 \ihmpc$, whereas the configuration-space FWHM scales as the inverse of the Fourier result and for our case, it is about ${\rm FWHM} \approx 4.2 \mpcoh$. We see some ringing in the lower panel, but its amplitude is much smaller compared with the peak. The horizontal green line denotes the FWHM of the window function in real and Fourier space.} 
    \label{fig:W}
\end{figure}

The final step involves converting the three-dimensional quantity, $\hat{P}^{\ast, tt}(\bk)$, into a correlation function by applying an inverse Fourier transform:
\begin{equation}
    \hat{\xi}^{\ast, tt}(\mathbf{r}) = {\rm IFT}[\hat{P}^{\ast, tt}(\bk)]
\end{equation}
and then weighting the derived three-dimensional correlation function by the appropriate Legendre polynomials and binning in $r$ to arrive at $\hat{\xi}^{\ast, tt}_\ell(r)$. We note that due to the finite mesh size in the power spectrum calculations ($N_{\rm grid} \sim 10^3$), on small scales the correlation function suffers from the effect of ``ringing,'' which we ameliorate in two ways. First, we apodize the three-dimensional power spectrum grid with a modified Blackman-Harris window\footnote{For other examples of apodization filters and their Fourier transforms, see \href{https://en.wikipedia.org/wiki/Window_function}{https://en.wikipedia.org/wiki/Window\_function}.} in each dimension:
\begin{eqnarray}
\label{eq:apod}
W(x) = 
\begin{cases}
    1 & k < b \\
    a_0 + a_1 \cos(x) + \\ a_2 \cos(2x) + a_3 \cos(3x)
    & b < k < b + a \\
    0 & k > b + a
\end{cases} \;
\end{eqnarray}
where $x \equiv \pi(k-b)/a$, $a = (k_{\rm Ny}/2)/0.68636$, $b = k_{\rm Ny}/4$, and $k_{\rm Ny} \equiv \pi N_{\rm mesh}/L_{\rm box}$ is the Nyquist frequency. Here, we have made use of the fact that the full width at half maximum (FWHM) for the Blackman-Harris window function is ${\rm FWHM} \approx 0.68636\, a$, so for our settings, we choose $a$ such that ${\rm FWHM} = k_{\rm Ny} \approx 1.8 \ihmpc$. In configuration space, FWHM scales as the inverse of the Fourier result, which corresponds to roughly ${\rm FWHM} \approx 4.2 \mpcoh$. We show the tapering function in Fig.~\ref{fig:W} both in Fourier and in configuration space for the parameters specified above. There is some vestigial ringing in configuration space, though its amplitude is diminished compared with the peak. We have also done several tests on smooth theoretical power spectra to asses the effect of applying our modified tapering function and have found that the effect around the BAO scale is negligible ($\sim$0.1\%). We additionally tried a cosine tapering and found it to perform similarly (albeit marginally worse) to that of our modified Blackman-Harris window.

\begin{figure}
    \centering
    \includegraphics[width=0.48\textwidth]{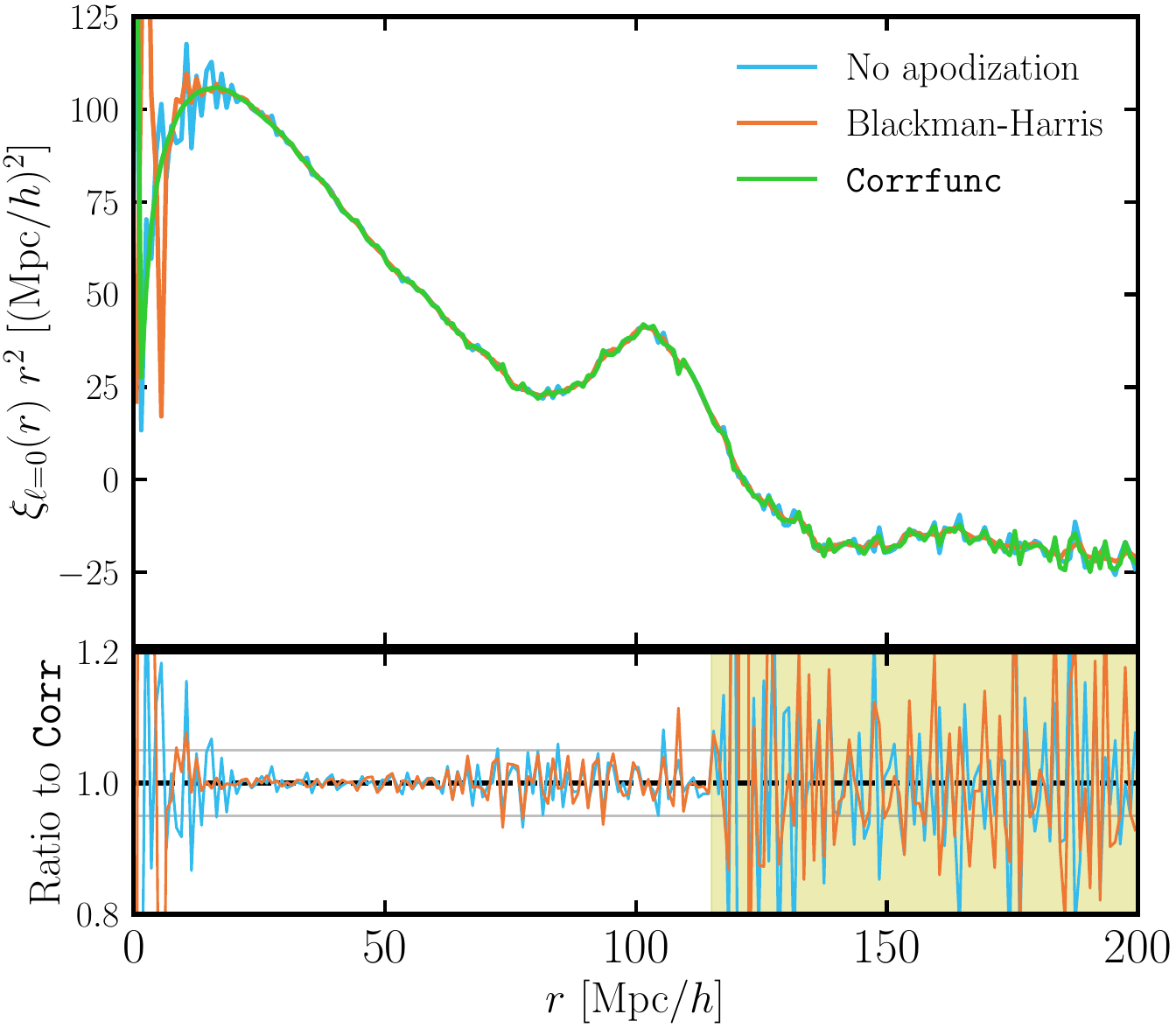} 
    \caption{Correlation function monopole computed using the inverse-Fourier transform with and without apodization (orange, blue), and direct pair-counting via \texttt{Corrfunc} (green). In the case where no tapering is applied, the ringing effects are evident across a wide range of scales, whereas the apodization of Eq.~\ref{eq:apod} mitigates those. On small scales, $r < 15 \mpcoh$, there are still clear boundary effects, which can be handled by supplying direct pair counts, as they are smooth in that regime. On large scales, the noise is large both for the pair counting and for the inverse-Fourier transform. We have demarcated this `noisy' region in the lower panel by a yellow band. The \texttt{Corrfunc} measurements are computed using the raw galaxy catalogs.}
    \label{fig:xi_apod}
\end{figure}

Second, we transition to the direct (brute-force) pair counting result on small scales of the \textbf{raw} (i.e., before applying CV) catalogs:
\begin{equation}
    \label{eq:comb}
        \xi^{\rm comb}_\ell(r)=\left[ 1-w(r)\right] \hat{\xi}^{\ast, tt}_\ell(r)+w(r)\,\hat{\xi}^{\texttt{Corr}, tt}(r),
\end{equation}
where $\hat{\xi}^{\texttt{Corr}, tt}(r)$ is the pair-count output of the natural estimator, $DD(r)/RR(r) - 1$, with the data-data pairs being computed from the raw catalogs via \texttt{Corrfunc}, and $w(r)$ is the weighting function, given by
\begin{equation}
    w(r)\equiv \frac{1}{2}\left[1-{\rm tanh}\left(\frac{r-r_{\rm pivot}}{\Delta r_w}\right)\right],
\end{equation}
where we set $r_{\rm pivot} = 2 \pi / (k_{\rm Ny}/4)$ and $\Delta r_w = 2 \pi / (k_{\rm Ny}/2)$, which ensures smooth interpolation between the two limits. Given the box size, $L_{\rm box} = 2\gpcoh$ and the grid size we adopt when measuring the power spectrum, $N_{\rm mesh} = 1152$, this corresponds to $r_{\rm pivot} = 13.9 \mpcoh$, $\Delta r_w = 6.9 \mpcoh$. We demonstrate in Fig.~\ref{fig:xi_apod} that these settings curtail some of the ringing effects. In particular, the curve corresponding to the non-apodized inverse-Fourier-transform displays ringing over a wide range of scales due to the sharp edges around $k_{\rm Ny}$. On the other hand, the brute-force pair counting (with \texttt{Corrfunc}) yields much smoother behavior across these scales albeit at a higher computational cost. Finally, the apodized result appears to have much better behavior over a wide range of scales compared with the case of no apodization and uses fewer resources than \texttt{Corrfunc}. On small scales, $r < r_{\rm pivot}$, there is still some vestigial ringing, which justifies combining the inverse-Fourier-transform result with pair counting. To decrease the computational demand of this operation, we can compute the pair counts only up to a few times the pivot scale, or alternatively, randomly downsample the galaxy sample used in the computation, as that would leave the small-scale correlation function almost unchanged. In addition, on very large scales, we still see a fair amount of noise. One can produce smoother curves by supplying a theoretical prediction of the clustering on these ultra-large scales.

\section{Post-reconstruction control variates}
\label{sec:post}

In this Section, we provide a brief overview of the standard reconstruction methods used in BAO analysis and develop the formalism of linear control variates (LCV) in the context of post-reconstruction density fields.  While it is certainly possible to develop a Zeldovich model for post-reconstruction fields \citep[see e.g.,][]{2019JCAP...09..017C}, we opt to use a linear model instead.  We believe the advantage of developing a more sophisticated procedure would be insignificant on the scales of interest to this study (i.e., the largest scales used in BAO analysis). The main reason for adopting the ZA instead of linear theory when applied to the pre-reconstruction galaxy catalogs is that much of the decorrelation between initial and final densities comes from the large-scale displacements, which the ZA models well, rather than e.g., the growth of structure. However, reconstruction aims to remove precisely these displacements, which also removes many of the advantages of ZA over linear theory in this case.  We thus expect LCV to provide a very good approximation for the post-reconstruction samples on large scales (as we demonstrate below).

\subsection{Reconstruction formalism}
\label{sec:recon}

The standard reconstruction procedure, applied to some tracer such as galaxies, halos, or particles within our N-body simulation, consists of the following steps \citep{ESSS07}:
\begin{enumerate}
\item The tracer density field $\delta_g$ is smoothed with a low-pass filter $\mathcal{S}$ that removes the small-scale signal. A typical choice for $\mathcal{S}$ is the Gaussian, $\mathcal{S}(k) =  \exp[-(kR_s)^2/2]$, of some smoothing scale $R_s$ (usually of order $\sim$10$\mpcoh$). 
\item Next, the shift, $\mathbf{\chi}$, is computed by unbiasing the smoothed galaxy density field (i.e., dividing by the linear bias and the linear redshift-space factor) and then taking the inverse gradient. In the periodic box this is equivalent to:
\begin{equation}
  \mathbf{\chi}_{\mathbf{k}} = -\frac{i\mathbf{k}}{k^2}
  \mathcal{S}(k)\ \Big( \frac{\delta_g(\mathbf{k})}{b + f\mu^2} \Big),
\label{eqn:recon_shift}
\end{equation}
where the linear bias is related to the Lagrangian first-order bias by $b=1+b_1$ and the line-of-sight angle $\mu = \hat{n} \cdot \hat{k}$. We note that in the limit of very large scales, where the approximations of scale-independent bias and supercluster infall hold, the calculated shift field approaches the negative smoothed Zeldovich displacement, i.e.\  $\mathbf{\chi}_{\mathbf{k}} \approx - \mathcal{S}(k) \bPsi^{(1)}(\bk)$.
\item The tracers are then moved by $\chi_d = \mathbf{R} \chi$ to form the ``displaced'' density field, $\delta_d$, where the matrix $\mathbf{R}$ is defined in Eq.~\ref{eq:zspace_za_displ} in the plane-parallel approximation.
\item For the ``randoms,'' an initially spatially uniform distribution of particles is shifted by
\begin{itemize}
    \item \textbf{RecSym}: $\chi_s = \mathbf{R}\chi,$ 
    \item \textbf{RecIso}: $\chi_s = \chi$,
\end{itemize}
depending on the method, and the ``shifted'' density field, $\delta_s$, is calculated \citep[e.g.,][]{Padmanabhan12,2015MNRAS.450.3822W,2016MNRAS.457.2068C,2016MNRAS.460.2453S,2018MNRAS.479.1021D,2019JCAP...09..017C}. The naming convention comes from the fact that the \textbf{RecIso} convention `isotropizes' the reconstructed field on large scales, whereas \textbf{RecSym} treats symmetrically $\delta_d$ and $\delta_s$. For a more thorough review on these reconstruction methods and the systematics associated with them, we refer the reader to \citet{Chen2023} and \citet{ChenDingPaillas2023}.
\item The reconstructed density field is finally defined as $\delta_r\equiv \delta_d-\delta_s$, and its power spectrum is obtained as $P_{\rm recon}(k) = P^{dd} + P^{ss} - 2 P^{ds} \propto \langle \left| \delta_r^2\right|\rangle$.
\end{enumerate}
We note that in real space, $f = 0$, and \textbf{RecSym} and \textbf{RecIso} become equivalent. In addition, ``displaced'' tracer has the same bias functional as the original tracer field, i.e., $F^d \equiv F^g$, while the ``shifted'' tracer is unbiased, i.e., $F^s\equiv 1$. For details on the reconstruction settings used in this work, we refer the reader to \citet{Fernandez2023} and \citet{Quintero2023}.

\subsection{Theoretical model}
\label{sec:recon_theory}

In order to develop the LCV formalism for mitigating the noise on the power spectrum and correlation function of a reconstructed field, we first need to obtain the analytic predictions for the reconstructed power spectrum mean.

In real space, we can express the auto- and cross-spectra of the displaced and shifted fields as follows \citep[see e.g.,][]{2019JCAP...09..017C}:
\begin{align}
    &P^{dd}_L(k) = \left[1 - \mathcal{S}(k)\right]^2 P_L(k), \\
    &P^{ds}_L(k) = -\mathcal{S}(k) \left[1 - \mathcal{S}(k)\right] P_L(k), \\
    &P^{ss}_L(k) = \mathcal{S}(k)^2 P_L(k) .
\end{align}
We note that this expression is equivalent for both \textbf{RecSym} and \textbf{RecIso}. In redshift-space, the power spectrum differs between the two and is given by:
\begin{align}
    \label{eq:sym}
    &P_{\rm sym}(\bk) = (b+f\mu^2)^2 P_{\rm L}(k) + \mathcal{O}(P_{\rm L}^2) \\
    &P_{\rm iso}(\bk) = \Big[(b+f\mu^2) (1 - \mathcal{S}) + b\ \mathcal{S} \Big]^2 P_{\rm L}(k) + \mathcal{O}(P_{\rm L}^2).
    \label{eq:iso}
\end{align}
As can be seen from these equations, \textbf{RecSym} restores supercluster infall at linear order, whereas \textbf{RecIso} diminishes the redshift-space distortions on large scales, but keeps them on small scales.

\subsection{Linear control variates}
\label{sec:lcv}

\begin{figure*}
    \centering
    \includegraphics[width=0.98\textwidth]{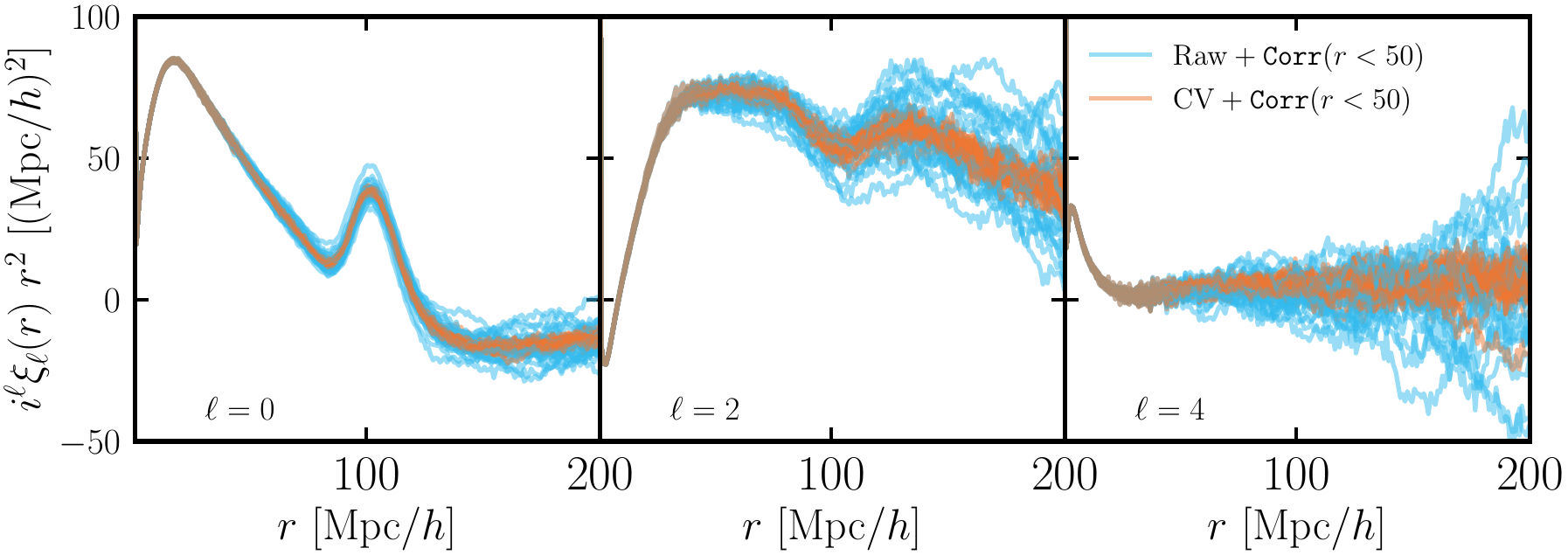} 
    \caption{Demonstration of the LCV technique in configuration space applied to a simulated, DESI-like, galaxy sample. Each column corresponds to a correlation function multipole ($\ell = 0,\, 2,\, 4$). On small scales, we provide the direct pair-counting result following Eq.~\ref{eq:comb}. In blue, we show the direct result from applying reconstruction to each of the 25 \textsc{AbacusSummit} simulations. In orange, we show the reconstructed correlation functions using the control variates method outlined in Section~\ref{sec:lcv}. Evidently, the CV-reduced curves have much less scatter compared with the raw outputs while retaining an unbiased mean, providing visual validation of the LCV technique in configuration space.} 
    \label{fig:xi_lcv}
\end{figure*}

Armed with the analytical expressions for reconstruction, we can write down the LCV equation:
\begin{equation}
    \hat{P}^{\ast, rr}_{\ell}(k) = \hat{P}^{rr}_{\ell}(k) - \beta_{\ell}(k) \left(\hat{P}^{LL}_{\ell}(k) - P^{LL}_{\ell}(k)\right) ,
\end{equation}
where $\hat{P}^{rr}_{\ell}(k)$ is the measured reconstructed power spectrum for a given tracer, whereas $P^{LL}_{\ell}(k)$ and $\hat{P}^{LL}_{\ell}(k)$ are the analytical and measured reconstructed power spectra for that tracer modeled using linear theory (see Eq.~\ref{eq:sym} and Eq.~\ref{eq:iso} depending on the reconstruction method). As before, we adopt the disconnected approximation for $\beta_\ell(k)$:
\begin{equation}
    \beta_{\ell}(k) = \left[ \frac{\hat{P}^{rL}_{\ell}(k)}{\hat{P}^{LL}_{\ell}(k)} \right]^2,
\end{equation}
where $\hat{P}^{rL}$ is the measured cross-power spectrum between the true and the modeled reconstructed fields.

As in the case of ZCV, $\beta_\ell(k)$ is assumed to be uncorrelated with the $X$ and $C$ variables (see Eq.~\ref{eq:cv}), so analogously to ZCV, we apply damping on small scales and smoothing with a Savitsky-Golay filter. Because the linear approximation is less accurate than the ZA, we opt to fit for the bias, $b$, only up to $k < 0.08 \ihmpc$.

To make an analytical prediction of the reconstructed power spectrum, we use the CLASS-generated \citep{2011arXiv1104.2932L} linear power spectrum for the \textsc{AbacusSummit} initial conditions and compute the theory-predicted multipoles, truncating at the hexadecapole, $\ell = 4$,
\begin{equation}
    P^{LL}_{\ell}(k) = b^2 C_{\ell}(\tilde \beta) P_L(k)
    \label{eq:p_ell}
\end{equation}
where
\begin{eqnarray}
C_{\ell}(\tilde \beta) &\equiv& \frac{2\ell+1}{2}\int_{-1}^{1} d\mu\, \left(1 + \tilde \beta \mu^2\right)^2 \mathcal{L}_{\ell}(\mu) \nonumber \\
&=&
\begin{cases}
    1 + \frac{2}{3}\tilde \beta + \frac{1}{5}\tilde \beta^2 & \ell = 0 \\
    \frac{4}{3}\tilde \beta + \frac{4}{7}\tilde \beta^2 & \ell = 2 \\
    \frac{8}{35}\tilde \beta^2 & \ell = 4
\end{cases} \; .
\label{eq:c_ell}
\end{eqnarray}
We define $\tilde \beta \equiv f_{\rm eff}/b$; for \textbf{RecSym}, $f_{\rm eff} = f$ while for \textbf{RecIso}, $f_{\rm eff}(k) = f [1-\mathcal{S}(k)]$. Note that in the latter case, $\tilde \beta$ is a function of $k$. 

Finally, we turn our attention to the correlation function multipoles. As in the case of ZCV (see Eq.~\ref{eq:zcv_xi}), we work with three-dimensional quantities, and express the LCV-reduced power spectrum measurement as follows:
\begin{equation}
    \hat{P}^{\ast, rr}(\bk) = \hat{P}^{rr}(\bk) - \beta(\bk) \left(\hat{P}^{LL}(\bk) - P^{LL}(\bk)\right) ,
\end{equation}
where $\beta(\bk)$ and $P^{LL}(\bk)$ are expanded into the three-dimensional $\bk$-grid from their multipole counterparts (see the discussion around Eqs.~\ref{eq:beta_leg} and \ref{eq:ZZ_leg}). We then apply an inverse Fourier transform to the LCV-reduced three-dimensional field
\begin{equation}
    \hat{\xi}^{\ast, rr}(\mathbf{r}) = {\rm IFT}[\hat{P}^{\ast, rr}(\bk)]
\end{equation}
and bin into Legendre multipoles, $\xi_\ell(r)$. On small scales, we supply direct pair counts using \texttt{Corrfunc} (analogously to Eq.~\ref{eq:comb}). We note that the correlation function estimator in that case is defined as $\xi(r) = [DD(r) - 2 DS(r) + SS(r)]/RR(r) - 1$, where $DD$, $DS$, $SS$ and $RR$ denote the data-data, data-shifted, shifted-shifted and random-random pairs, respectively. The `data' here corresponds to the \textbf{raw} `shifted' galaxy catalogs, i.e., before LCV reduction is applied to them. As a showcase, we present a visual demonstration of the LCV-reduced reconstructed correlation functions in Fig.~\ref{fig:xi_lcv} for a DESI-like galaxy sample (see Section~\ref{sec:app} for more details regarding the construction of the mock catalogs). We discuss the details of the CV-induced reduction in the next section. As a result of applying LCV, the scatter across the 25 boxes is evidently diminished while the mean appears unbiased.

\section{Performance in DESI samples}
\label{sec:app}

In this Section, we summarize our findings from applying the ZCV and LCV methods to realistic mocks of the DESI survey.

\subsection{Dark Energy Spectroscopic Instrument}
\label{sec:desi}

The main impetus for this work is the need to robustly test the analysis pipeline of the DESI redshift survey in anticipation of forthcoming early data analysis within a limited computing budget.

DESI is a Stage IV dark energy experiment currently conducting a five-year survey of about a third of the sky with the goal to amass spectra for approximately 40 million galaxies and quasars \citep{2016arXiv161100036D}. The instrument operates on the Mayall 4-meter telescope at Kitt Peak National Observatory \citep{2022AJ....164..207D} and can obtain simultaneous spectra of almost 5000 objects over a $\sim$3$^{\circ}$ field \citep{2016arXiv161100037D,2023AJ....165....9S} thanks to a robotic, fiber-fed, highly multiplexed spectroscopic instrument. The goal of the experiment is to unravel the nature of dark energy through precise measurements of the expansion history \citep{2013arXiv1308.0847L} and thus the dark energy equation of state parameters $w_0$ and $w_a$, with a predicted factor of five to ten improvement on their error relative to previous Stage-III experiments \citep{2016arXiv161100036D}. Additionally, the redshift clustering of galaxies will provide a window into the growth-of-structure of the Universe and allow us to constrain the parameter combination, $f\sigma_8$.

\subsection{\textsc{AbacusSummit}}
\label{sec:abacus}

\textsc{AbacusSummit} is a suite of cosmological $N$-body simulations designed to meet and exceed the Cosmological Simulation Requirements of the DESI survey \citep{2021MNRAS.508.4017M}. The simulations were run with \textsc{Abacus} \citep{2019MNRAS.485.3370G,2021MNRAS.508..575G}, a high-accuracy, high-performance cosmological $N$-body simulation code, optimized for GPU architectures and  for large-volume simulations, on the Summit supercomputer at the Oak Ridge Leadership Computing Facility.

The majority of the \textsc{AbacusSummit} simulations are made up of the \texttt{base} resolution boxes, which house 6912$^3$ particles in a $2\gpcoh$ box, each with a mass of $M_{\rm part} = 2.1 \times 10^9\msunoh$. While the \textsc{AbacusSummit} suite spans a wide range of cosmologies, here we focus on the fiducial outputs (\textit{Planck} 2018: $\Omega_b h^2 = 0.02237$, $\Omega_c h^2 = 0.12$, $h = 0.6736$, $10^9 A_s = 2.0830$, $n_s = 0.9649$, $w_0 = -1$, $w_a = 0$). In particular, we employ the 25 \texttt{base} boxes \texttt{AbacusSummit\_base\_c000\_ph\{000-024\}} and utilize the halo and particle catalogs \citep{2022MNRAS.509..501H} as well as initial conditions outputs. For full details on all data products, see \citet{2021MNRAS.508.4017M}.

\subsection{A measure of success}
\label{sec:rhoxc}

The cross-correlation coefficient between the modeled tracer power spectrum, be it through the ZA or linear theory, and the measured tracer power spectra quantifies the effectiveness of the control variates technique. We compute this quantity as
\begin{align}
    \rho_{xc} &= \frac{\textrm{Cov}[\hat{P}^{tt}_{\ell}(k), \hat{P}^{ZZ}_{\ell}(k)]}{\sqrt{\textrm{Var}[\hat{P}^{tt}_{\ell}(k)]\textrm{Var}[\hat{P}^{ZZ}_{\ell}(k)]}}\,.
    \label{eq:rhoxc}
\end{align}
\noindent 
The above expression applies to the ZCV case; in the LCV case, we can simply swap the tracer with the reconstructed field, $t \rightarrow r$, and the ZA with the linear prediction, $Z \rightarrow L$. The fractional variance reduction purveyed by the CV method is equal to $1-\rho_{xc}^2$, making this quantity of central interest to this technique. Similarly to the case of $\beta_\ell(k)$, we employ the disconnected approximation when estimating $\rho_{xc}$, which has been shown to hold to very high accuracy for $k<0.2\ihmpc$ \citep{2020PhRvD.102l3517W}.

It is important to emphasize that both the shot noise of the sample and its satellite fraction play an important role in decorrelating the $X$ and $C$ random variables and hence reduce the efficacy of the control variates technique. Reassuringly, \citet{DeRose23} find that the ZCV method saturates the shot-noise limit of $\rho_{xc}$, suggesting that the ZCV technique is optimal in reducing the noise given the unavoidable limitation of sample shot noise. Similarly, we find that $\rho_{xc}$ derived from LCV recovers the shot-noise limit on large scales, though is slightly suboptimal on intermediate scales $k \sim 0.4 \ihmpc$. We attribute this to the limitation of modeling the reconstructed power through linear theory and leave a higher-order modeling for future work. We stress, however, that the most substantial gains that are also of interest to BAO and large-scale-structure science come from the largest scales, for which the method is optimal. We note that in the above calculations, we opt not to subtract shot noise from the measurements so as to give the user freedom to adopt a shot noise model of their choice. 

\subsection{Galaxy mock catalogs}
\label{sec:mock}

Before BAO analysis can be performed with confidence on real data, there are a number of systematics checks that need to be done on synthetic galaxy mocks mimicking the galaxy populations targeted by the DESI survey. One of the immediate applications of the CV method is the reduction of the measurement noise on these mocks, which enables the testing of the analysis pipeline in the regime of sub-percentage precision. Such precision will be achieved by the DESI Year 3 (Y3) and Year 5 (Y5) datasets, but cannot be reached by the \textsc{AbacusSummit} suite alone.

Here, we apply our CV formalism to two of the populations that DESI is utilizing for cosmology: luminous red galaxies (LRGs) \citep{Fernandez2023} and emission-line galaxies (ELGs) \citep{Quintero2023}. Tens of mock catalogs have been generated for each tracer using various extensions of the standard HOD method for all 25 of the fiducial cosmology simulations. The majority of ELG catalogs are at $z = 1.1$ (with a couple at $z = 0.8$), while all the LRG ones are at $z = 0.8$. We will present the results from a single HOD (the `main HOD' or `first-generation mocks' of \citealt{Alam2023,Fernandez2023,Quintero2023}).  We find similar behavior for the other mock catalogs. In the near term, we plan to extend the application of the CV technique to mock catalogs of the population of DESI quasi-stellar objects (QSOs).

In the subsequent sections, we focus on the two-point statistics, $P_\ell(k)$ and $\xi_\ell(k)$, calculated from the LRG and ELG synthetic catalogs generated as part of the first-generation mocks. We utilize both `pre-' and `post-reconstruction' samples. `Pre-reconstruction' 
in this context simply refers to the raw mock galaxy outputs at a given redshift.

\subsubsection{Luminous red galaxies}

\begin{figure*}
    \centering
    \includegraphics[width=0.98\textwidth]{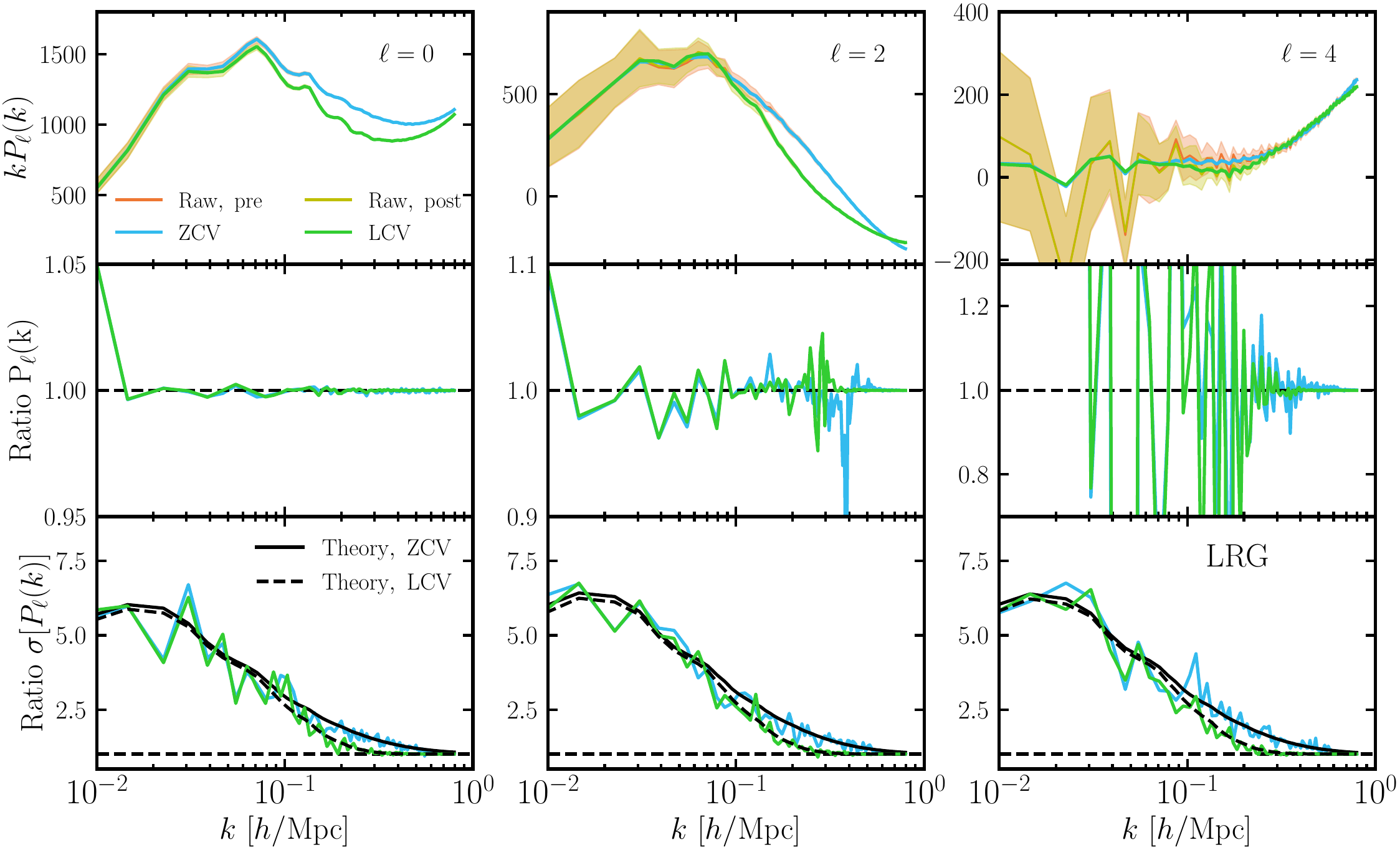} 
    \caption{Reduction of the noise in the measured power spectrum of DESI-like LRGs using Zeldovich control variates (ZCV) and Linear control variates (LCV) from all 25 fiducial \textsc{AbacusSummit} boxes. Each column corresponds to a Legendre multipole ($\ell = 0$, 2, 4). The top row shows the raw and the CV-reduced power spectrum multipoles. The orange and yellow bands denote the error on the raw measurements before and after applying reconstruction, which undoes the non-linear effects to first order (see Section~\ref{sec:recon}). The middle row shows the ratio between the means of the raw and the CV multipoles. The agreement between the two is excellent for $\ell = 0$ and $\ell = 2$, and dominated by noise for $\ell = 4$. On very large scales, we see deviations due to sample variance. We check this by studying the ratio between the ZCV (LCV) prediction from theory and simulation and find that the noise imprint is virtually indistinguishable from the ratio shown in the middle panel. The bottom row shows the reduction of noise yielded by the ZCV (LCV) technique on this sample. We find good agreement between the numerical and the theoretical reduction in noise (see Eq.~\ref{eq:rhoxc}). On large scales, the error on the multipoles is reduced by a factor of 6, with $\ell = 2$ and  $\ell = 4$ having a slightly more improved measurement compared with $\ell = 0$. The gain from using LCV is similar to ZCV (a factor of 6 reduction on large scales). We note, however, that past $k > 0.1 \ihmpc$ the correlation coefficient decays faster, yielding a smaller error bar shrinkage compared with ZCV.}
    \label{fig:pk_lrg}
\end{figure*}

The first tracer we will look at is luminous red galaxies (LRGs), which are selected effectively by applying a magnitude limit as a function of color, which allows the most luminous objects at a given redshift to be selected. The number of LRGs in the first-generation mock catalogs at $z = 0.8$ is on average a bit over 8 million, corresponding to a comoving number density of about $\bar{n}_{\rm LRG} = 1.0 \times 10^{-3}\, h^3 {\rm Mpc}^{-3}$, while their linear bias is roughly $b_{\rm LRG} \approx 2.4$. This is similar, though slightly higher, than the number density of observed galaxies at that redshift. In this Section, we explicitly show the effect of applying LCV and ZCV only on the LRG power spectrum multipoles and show the correlation function results in App.~\ref{app:add}, which are qualitatively similar. Furthermore, we note that when adopting the CV method to obtain tighter constraints on the BAO parameters, \citet{Fernandez2023} find consistency between the gained signal-to-noise in the power spectrum and the correlation function.

In Fig.~\ref{fig:pk_lrg}, we present both the ZCV- and LCV-reduced power spectrum measurements from all 25 fiducial \textsc{AbacusSummit} boxes for the three Legendre multipoles ($\ell = 0$, 2, 4). Starting with ZCV, We see that the agreement between the raw and the CV outputs is excellent for $\ell = 0$ and $\ell = 2$ (within 0.3\% and 3\%, respectively), whereas we have checked that the differences for all multipoles (including $\ell = 4$) are consistent with noise. The extra numerical noise in the ratio plot for $\ell = 2$ at $k \approx 0.4 \ihmpc$ is due to the power spectrum crossing zero. Reassuringly, we find good agreement between the numerically computed reduction in noise from the samples and the theoretical prediction coming from the cross-correlation coefficient, $\rho_{xc}$ (see Eq.~\ref{eq:rhoxc}). In particular, on large scales, we see that the noise for all multipoles is reduced by a factor of 6, with $\ell = 2$ and  $\ell = 4$ yielding slightly higher signal-to-noise than $\ell = 0$. This scale dependence is expected given that the goal of the Zeldovich approximation is to capture the first-order large-scale displacements.

Fig.~\ref{fig:pk_lrg} also illustrates the effect of the LCV method on the LRG catalogs. We note that while the LRG HOD is identical both post- and pre-reconstruction, we undo the non-linear effects to first order for the post-reconstruction catalogs (see Section~\ref{sec:recon}). Reassuringly, similarly to the ZCV case, we do not see biases in the CV-reduced power spectra, and the gain in applying LCV is similar: the measurement noise is reduced by a factor of 6 on large scales. We note, however, that past $k > 0.1 \ihmpc$ the correlation coefficient decays faster than in the ZCV case, yielding a smaller improvement in the measurement noise.

When comparing the results of applying CV on the correlation function using the ZA and the linear approximation, we notice a similar trend (see Fig.~\ref{fig:xi_lrg}). Namely, the CV-reduced curves do not appear to be biased relative to the raw outputs, though the $\ell = 4$ curves receive a substantial noise contribution due to sample variance. Both ZCV and LCV gain us a factor of 4 in noise reduction for $\ell = 0$ and $\ell = 2$ and slightly less for $\ell = 4$. As we will see in the next Section on ELGs, the CV-induced noise mitigation is much less scale-dependent in configuration space compared with Fourier space due to mode coupling, which smears the large-scale reduction (i.e., small $k$) across the physical scales. We note that this does not imply that the reduction in noise of the correlation function is less than that of the power spectrum.

\subsubsection{Emission-line galaxies}

\begin{figure*}
    \centering
    \includegraphics[width=0.98\textwidth]{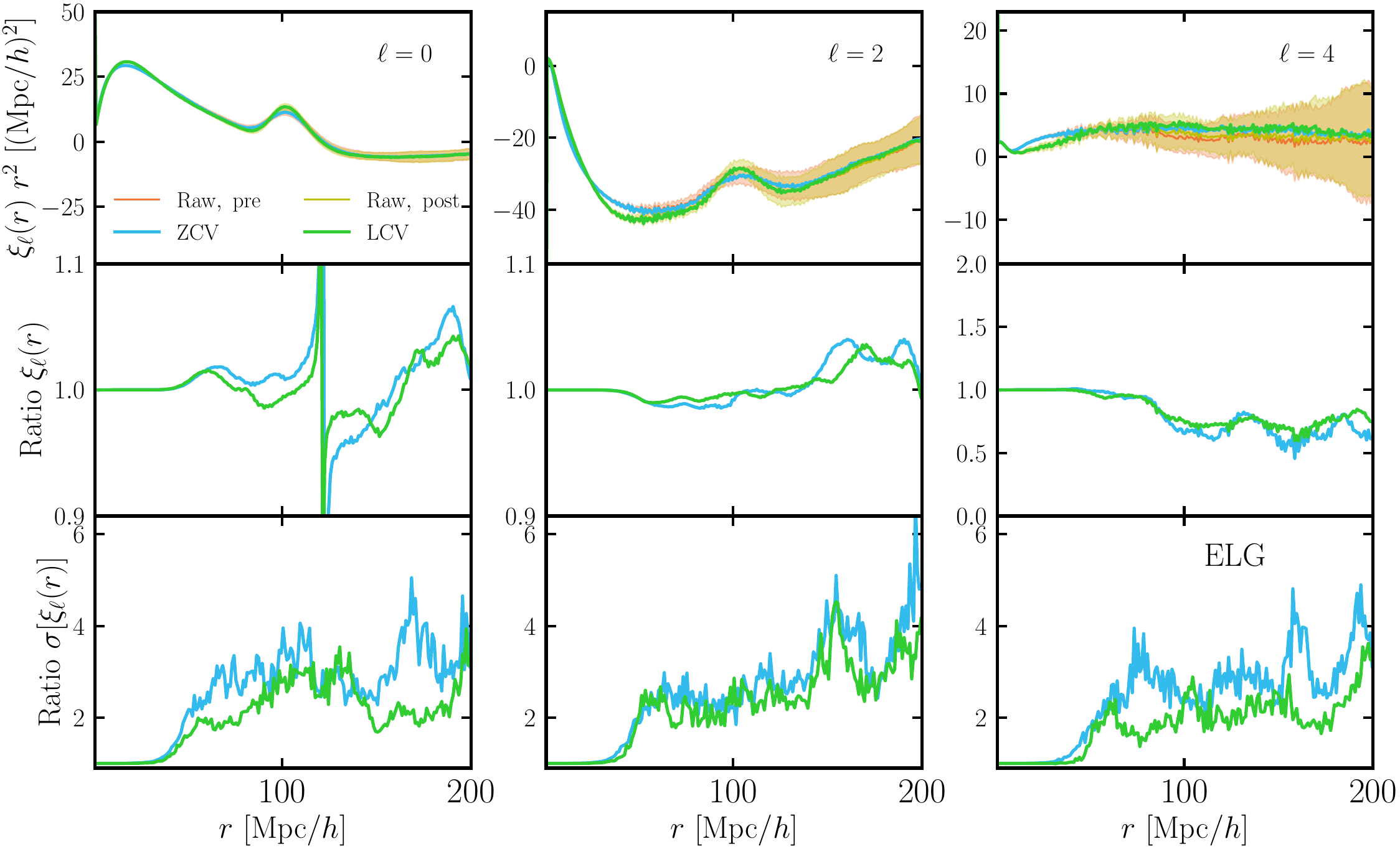} 
    \caption{
    Reduction of the noise in the measured correlation function of DESI-like ELGs using Zeldovich control variates (ZCV) and Linear control variates (LCV) from all 25 fiducial \textsc{AbacusSummit} boxes. The top row shows the raw and the CV-reduced correlation function multipoles along with the error in the raw measurements (orange and yellow band). The middle row shows the ratio of the multipoles between the raw means and the CV-reduced means. Both $\ell = 0$ and $\ell = 2$ appear to be unbiased until $r \gtrsim 150 \mpcoh$, whereas $\ell = 4$ sees a larger deviation from one at smaller pair distances. As can be seen from the top row, this regime is dominated by sample variance effects, and we thus attribute this finding to volume limitations. This is also corroborated by the fact that the noise imprint of the ZCV curves is similar to that of the LCV curves. The bottom row shows that we gain a factor of 3 (2.5) in noise reduction by employing the ZCV (LCV) technique for $\ell = 0$ and $\ell = 2$ and slightly less for $\ell = 4$. Below $r \lesssim 50 \mpcoh$, we do not see any improvement in the signal-to-noise, as in that regime, we supply direct pair counts from \texttt{Corrfunc}.} 
    \label{fig:xi_elg}
\end{figure*}

We now turn our attention to the other galaxy tracer of interest to this study, ELGs, which make up the majority of galaxies that DESI will observe. These galaxies are  characterized by prominent [O{\sc ii}] and [O{\sc iii}] emission lines as well as other less prominent features such as [Ne{\sc iii}] and Fe {\sc ii}${}^\star$ emission lines. The number of ELGs in the mock catalogs of \citet{Alam2023} is a bit over 24 million on average, corresponding to a comoving number density of about $\bar{n}_{\rm ELG} = 3.0 \times 10^{-3}\, h^3\, {\rm Mpc}^{-3}$. These are star-forming galaxies that are not as massive as the LRGs, and their linear bias is lower (roughly $b_{\rm ELG} \approx 1.2$). The redshift of all samples is $z = 1.1$. We note that \citet{Quintero2023} uses a higher-number-density version of these mocks. Here, we opt to adopt the lower-density catalogs, obtained via random downsampling, as their number density is closer to the anticipated number density of DESI ELGs and thus better represents the improvement we expect from CV for the survey. We have checked that the higher-density sample yields a larger improvement of the measured correlation function, as expected.

Fig.~\ref{fig:xi_elg} shows both the ZCV- and LCV-reduced correlation function multipoles for the ELG samples. We explore the results in the 25 boxes for both the raw and the CV outputs and find that visually the correlation function curves for both are in good agreement. When comparing the ratios between the two means, the deviations from one are small relative to the error. A numerical feature can be seen in $\ell = 0$ due to the correlation function crossing zero. We note that the biases in the middle panel have the same shape in both the ZCV and the LCV case (see Fig.~\ref{fig:xi_elg}), strongly suggesting that their source is numerical rather than theoretical, as the LCV and ZCV frameworks are independent of each other, and their only commonality is the sample variance and the input power spectrum, which matches exactly the linear power spectrum initializing the simulation. We have checked through the $\chi^2$ statistics that these differences are consistent with noise. Compared with the LRGs (not shown), we see a smaller reduction in the noise across all scales when adopting the ZCV method: roughly a factor of 3 for the ELGs compared with 4 for the LRGs. As we noted in the previous section, this factor appears to be largely scale-independent in configuration space due to the mode coupling, which smears the large-scale reduction across a broader range of physical scales. Below $r < 50 \mpcoh$, we supply direct pair counts from \texttt{Corrfunc} and do not make use of the CV reduction on these scales. 

The ELG samples used in the LCV case are created by applying reconstruction to the standard mock outputs from Fig.~\ref{fig:xi_elg}. We find a consistent, though slightly lower, gain in signal-to-noise compared with the pre-reconstruction catalogs. This slight worsening of the CV performance could be attributed to the lower cross-correlation coefficient between the initial conditions field modeled with linear theory and the final reconstructed density field (see Fig.~\ref{fig:pk_lrg}). Generally, however, since reconstruction does not affect the bias and number density of the sample and at the same time, removes the first-order linear displacements, we expect the performance of the linear approximation to be extremely similar to the ZA, which is indeed what we observe.

Our findings for the effect of the CV method on the ELG power spectrum multipoles are qualitatively very similar to the LRGs, as can be seen in App.~\ref{app:add} and Fig.~\ref{fig:pk_elg}. The ratio of the CV-reduced and the raw power spectrum means appears to be unbiased, though $\ell = 4$ displays large oscillations around the mean. The predicted gain in precision from theory appears to be in good agreement with the measurement from simulations.
Compared with the LRGs, the increase in signal-to-noise on large scales is lower (a factor of 5 compared with a factor of 6). This is likely the case due to the higher satellite fraction and lower value of the combination $b^2 \bar{n}$, which implies a lower cross-correlation coefficient between the galaxies and the initial conditions fields.  The increase in precision is similar between ZCV and LCV, though as expected, the cross-correlation coefficient between the linear theory applied to the initial conditions and the final galaxy field drops down at lower $k$ values ($k \approx 0.1 \ihmpc$) than in the ZA case.

\subsection{Effect on BAO parameter constraints}
\label{sec:bao}

In this section, we discuss the performance of BAO parameter constraints with applications of ZCV and LCV in the aforementioned DESI mock catalogs. We focus on ELG mocks (see performance of various HOD mocks in \citet{Quintero2023}), but LRG mocks (see performance in \citet{Fernandez2023}) show similar trends. We first comment on the mean of the estimated BAO parameters, and then provide a quantitative evaluation for the reduction of the errors in the measurements, using an initial version of lower density ($\overline{n} \sim 3 \times 10^{-3} h^{3} {\rm Mpc}^{-3}$) ELG mocks (\citet[][]{Alam2023}).

We fit the BAO parameters \citep[e.g.,][a brief description of the details of the fits is below]{Beutler17} for each of the 25 mocks before and after applying CV (using the same covariance matrix), and obtain the mean and the error (estimated as the scatter among the 25 estimates) of the parameters. We constrain both the isotropic dilation and anisotropic warping parameters, $\{\alpha, \epsilon\}$ \citep{Padmanabhan08} (or equivalently the two scaling parameters, $\{\alpha_{\perp},\alpha_{\parallel}\}$ \citep{Anderson14}). We find that the mean of pre- and post-reconstruction constraints for $\{\alpha, \epsilon\}$ and $\{\alpha_{\perp},\alpha_{\parallel}\}$ with ZCV and LCV reduction, respectively, are consistent with the constraints using the raw measurements. As an example, for the Fourier space post-reconstruction analysis, the absolute differences between the CV and raw parameter values are 
$(0.002 \pm 0.001)$ for $\alpha$, 
$(0.002 \pm 0.002)$ for $\epsilon$, 
$(0.002 \pm 0.001)$ for $\alpha_{\perp}$
and $(0.006 \pm 0.004)$ for $\alpha_{\parallel}$, 
which is well within the error budget of the fits (0.005 to 0.01 across the four parameters) \citep[see][for a more in-depth discussion]{Fernandez2023,Quintero2023}. In configuration space, we see similar performance. These demonstrate that both ZCV and LCV approaches do not bias the BAO constraints.

\begin{figure}
    \centering
    \includegraphics[width=\columnwidth]{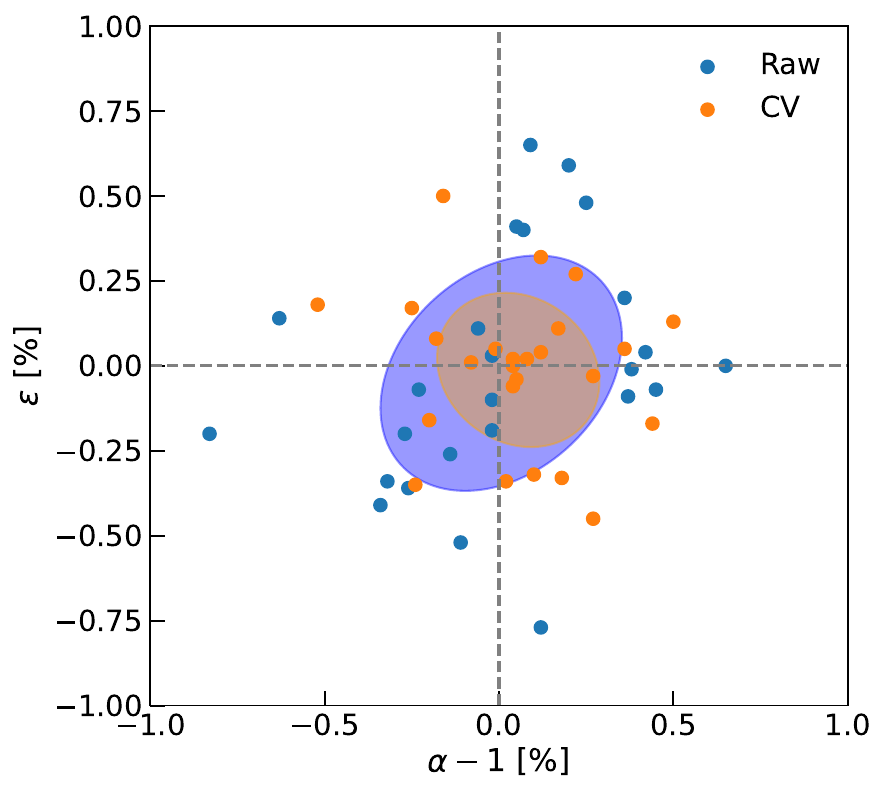}
    \caption{The 25 fits of $\{\alpha-1,\epsilon\}$ with raw data (blue) and with application of LCV (orange) for post-reconstruction power spectrum, shown as percentage. The two contours are 1 standard deviation covariance ellipses. Both the points and the contours show that the scatter among the 25 fits is reduced by CV, while the mean remains consistent. The reduction rate of the scatter is about 1.5 in both parameters. We note that both the raw and the CV fits use the same covariance matrix \citep{Variu2023}.}
    \label{fig:alpha-eps_data_scatter}
\end{figure}

We find that with CV reduction, the errors in the estimates of the BAO parameters are reduced both pre- and post-reconstruction. For example, in the post-reconstruction analysis in Fourier space, the reduction rates in errors (ratio of estimated errors pre- and post-CV) for $\alpha_{\perp}$, $\alpha_{\parallel}$, $\alpha$ and $\epsilon$ are 1.2, 1.8, 1.5 and 1.5, respectively (see Figure~\ref{fig:alpha-eps_data_scatter}). We obtain similar results in configuration space and pre-reconstruction. In the following, we predict the amount of reduction with LCV analytically and compare to what we see in the fits of the mocks. 

We assume a linear relation between the model and the parameters: $\boldsymbol{y} =\boldsymbol{y}_{0}+\sum_i  (\partial \boldsymbol{y} / \partial {{{\theta}}_{i}})\theta_{i}$, where  $\boldsymbol{y}$ is the model of the power spectrum (can also be correlation function), $\boldsymbol{y}_{0}$ is the best-fit model, and $\boldsymbol{\theta}$ is the parameter vector, which includes the BAO parameters and other physical as well as nuisance parameters. Fitting the data $\boldsymbol{d}$ with this model and minimizing $\chi^2$, we derive an expression that shows how a small change in the data vector $\delta \boldsymbol{d}$ propagates to a change in the estimated parameters $\delta\boldsymbol{\theta}$:
\begin{equation}\label{eq:M_mat}
    \delta \boldsymbol{\theta}=\boldsymbol{F}^{-1}\frac{\partial \boldsymbol{y}}{\partial \boldsymbol{\theta}}\boldsymbol{C}^{-1}\delta\boldsymbol{d} = \boldsymbol{\mathcal{M}}\ \delta \boldsymbol{d}.
\end{equation}
Here $\boldsymbol{F}$ is the Fisher matrix, $\boldsymbol{F}=\left(\partial \boldsymbol{y}/\partial \boldsymbol{\theta}\right)^{\intercal}\boldsymbol{C}^{-1}\left(\partial \boldsymbol{y}/\partial \boldsymbol{\theta}\right)$, and $\boldsymbol{C}$ is the covariance matrix used in the fits. Assuming the mean of the data is our best fit model and the deviation in individual mocks from the mean is small, we can use this expression to find how the scatter in data impacts BAO parameters in both raw and CV cases. We model the power spectrum following the template by \citet{ESW07} (see detailed discussion in Appendix C2 of \citealt{Chen22}) and with six broadband terms for each multipole.  We evaluate the derivatives at (1,0) for $\{\alpha,\epsilon\}$ and at the best-fit values for other physical parameters. We use the same raw data covariance matrix used in real data analysis, for both raw and CV cases, to be consistent with what is done in the mocks. 
This covariance matrix is evaluated from 1000 EZmocks 
\citep{2015MNRAS.446.2621C,Variu2023} and is independent of the cosmological parameters. In both mocks and forecasts, we use only the monopole and quadrupole power spectra and focus on the BAO parameters $\{\alpha, \epsilon\}$. 

We then randomly draw 1000 sets of 25 differential data vectors $\delta \boldsymbol{d}_{\rm raw}$ from the covariance matrix for the raw power spectra, where $\delta \boldsymbol{d}_{\rm raw}$ denotes the difference between the realization and the mean data vector. We rescale these data vectors by $\sqrt{1-\rho_{xc}^2}$ to get $\delta \boldsymbol{d}_{\rm cv}$. The corresponding changes in the parameters can then be obtained with Eq.~\ref{eq:M_mat}. Taking the variance of 25 $\delta \boldsymbol{\theta}$ for each of the 1000 trials gives a distribution of the error in $\delta \boldsymbol{\theta}$. The expectation of the errors in the parameters is given by 
\begin{equation}
    \left\langle\delta\boldsymbol{\theta}_{\rm raw}\, \delta {\boldsymbol{\theta}_{\rm raw}^\intercal}\right\rangle
    =\boldsymbol{\mathcal{M}}\ \boldsymbol{C} \boldsymbol{\mathcal{M}}^\intercal=\boldsymbol{F}^{-1}
\end{equation}
for the raw case, which agrees with the variance of a large number of $\delta \boldsymbol{\theta}_{\rm raw}$. For the CV case, we calculate the expectation by 
\begin{equation}
    \left\langle\delta \boldsymbol{\theta}_{\rm cv}\, \delta {\boldsymbol{\theta}_{\rm cv}^\intercal}\right\rangle
    =\boldsymbol{\mathcal{M}}\ \boldsymbol{C}_{\rm cv} \boldsymbol{\mathcal{M}}^\intercal,
\end{equation}
where $\boldsymbol{C}_{\rm cv}$ is the covariance matrix for $\boldsymbol{d}_{\rm cv}$, the data vector after the application of CV. We simply rescale the covariance matrix for the raw data by $\sqrt{(1-\rho_{xc}^2)_{i}(1-\rho_{xc}^2)_{j}}$, where $i$ and $j$ run through the $k$-binning indices of the two multipoles, to obtain the CV covariance matrix. We denote these expectations as our forecast.

\begin{figure*}
    \centering
    \includegraphics[width=\columnwidth]{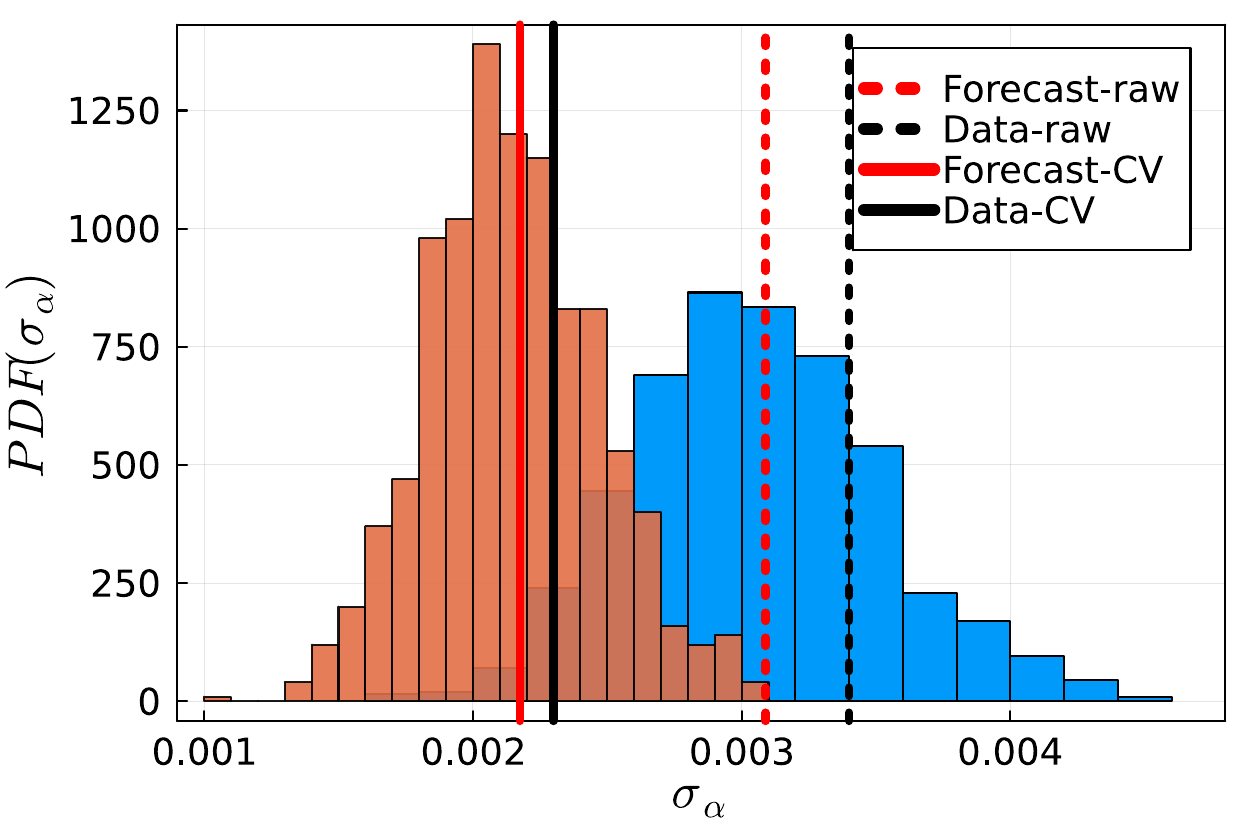}
    \includegraphics[width=\columnwidth]{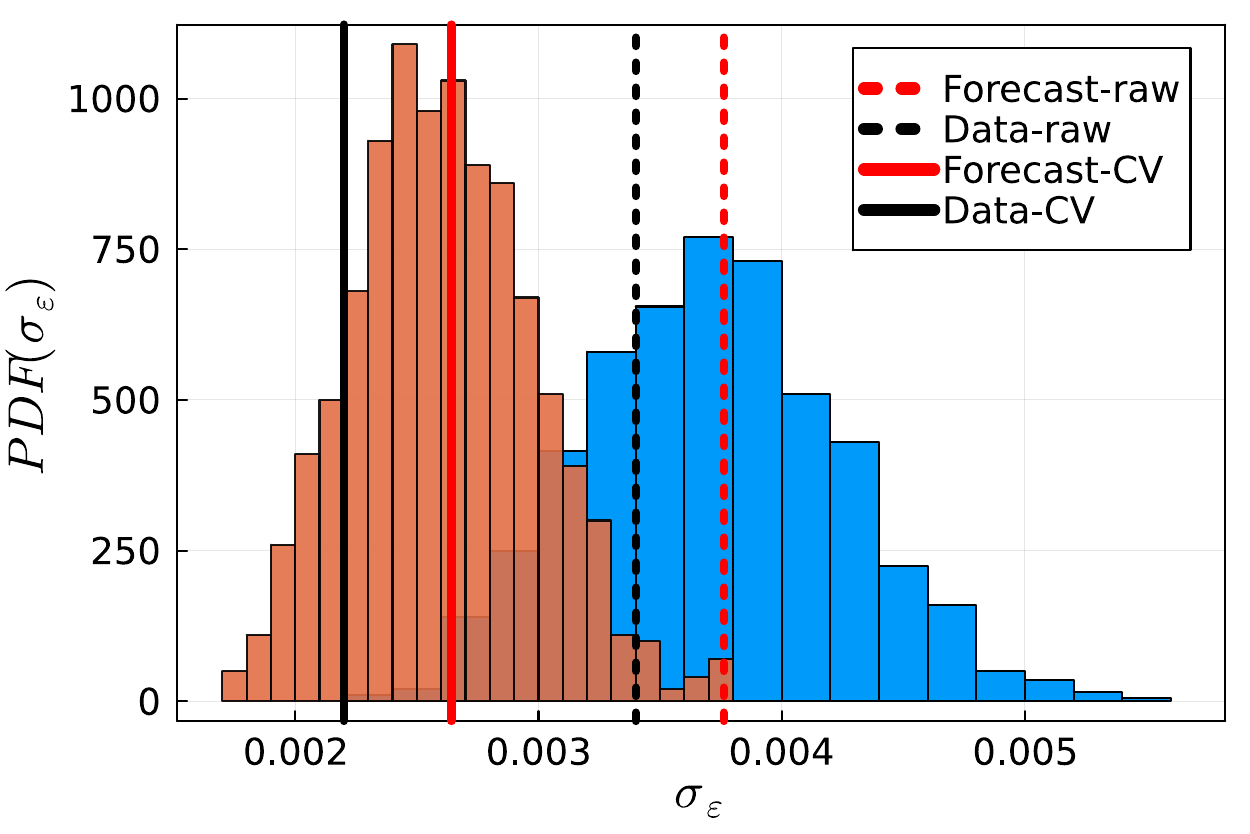}
    \caption{The distributions of the errors in  $\alpha$ (left) and $\epsilon$ (right) from 1000 random draws of 25 differential data vectors. The blue histograms show the resulting distributions of errors for the raw case. The red histograms show the CV counterpart. The vertical lines represent the expectations of the distributions (red) and results from the 25 \textsc{AbacusSummit} mocks (black) for both the raw (dashed) and CV (solid) cases. There are large spreads in the distributions of errors in both $\alpha$ and $\epsilon$, due to a relatively small number of mocks, but the scatters are lower in the CV case. The data results are consistent with our analytic predictions. }
    \label{fig:random_draw}
\end{figure*}

Fig.~\ref{fig:random_draw} shows the distributions of the estimated errors (standard deviation of each 25 $\delta\boldsymbol{\theta}$) in $\alpha$ and $\epsilon$ for the raw and CV cases from the 1000 trials, in comparison with the expectations and with the errors from real data fits. There is a large spread of the estimated errors from 1000 random draws, due to the relatively small number of mocks, although the spread goes down with CV. The errors from our single realization of 25 mocks are consistent with the distribution of the errors from analytic calculation. For $\alpha$, both the raw and CV cases have the error from data fits higher than but consistent with the forecast. The opposite occurs for $\epsilon$, so the estimated error in $\epsilon$ from the data is lower than but still consistent with our prediction. We have also tested that the reduction rates in $\alpha$ and $\epsilon$ are uncorrelated by drawing random samples from the covariance. In other words, even though our forecast and data fitting show roughly the same amount of reduction in both parameters, it would not be surprising if either $\alpha$ or $\epsilon$ error had a little more reduction than the other, for a new set of 25 mocks. 

The above discussion made a simple approximation for the post-control-variate covariance matrix. In particular, we scale the variances, but leave the correlation between different $k$ bins and multipoles the same. We defer a more careful construction of the covariance matrix for future work, but we found that our overall conclusions were insensitive to this particular approximation. In particular, we found that the CV measurements reduced the errors on the BAO distance parameters by a factor of $\sim 1.2-1.8$ . Achieving a similar reduction in errors would require a factor of $\sim 1.4$ to $3.2$ more simulation volume, emphasizing the benefit of this technique. Of course, the exact reduction in the error will depend on details of the particular galaxy sample, including its number density and bias.

\section{Conclusions}
\label{sec:conc}

In this paper, we apply the control variates formalism to realistic galaxy mock catalogs generated for the DESI survey with the goal of reducing the error on their two-point statistics in configuration and Fourier space. This allows us to generate low-noise, simulated data vectors for testing the robustness of the DESI analysis pipeline. The CV technique enables that without incurring the computational cost of running additional simulations and, as such, is an invaluable tool in the era of precision cosmological observations.

Below, we summarize the main points of the paper:
\begin{itemize}
    \item We provide a conceptual introduction to the control variates method alongside a simple visualization (see Fig.~\ref{fig:cv}) for building intuition. We then review the recently developed analytic approach of Zeldovich control variates (ZCV), and comment on its application to the \textsc{AbacusSummit} suite of simulations.
    \item In Section~\ref{sec:power2xi}, we extend the ZCV technique from Fourier space to configuration space via the correlation function multipoles and comment on some of the numerical challenges that arise. In Fig.~\ref{fig:xi_apod}, we demonstrate the effect of apodization with a Blackman-Harris filter (see Fig.~\ref{fig:W}) when performing an inverse Fourier transform to obtain the correlation function.
    \item In Section~\ref{sec:lcv}, we develop a related method for reducing the noise of two-point statistics computed with the `reconstructed' density fields, which we dub Linear control variates (LCV).  LCV takes advantage of the fact that density-field reconstruction removes much of the large-scale displacement that causes the initial and final density fields to decorrelate.  Since an improved modeling of this displacement is the major gain of ZCV over LCV, the latter is a pragmatic choice for post-reconstruction CV.  We demonstrate in Fig.~\ref{fig:xi_lcv} that our LCV outputs have greatly reduced scatter relative to the raw outputs.
    \item In Figs.~\ref{fig:pk_lrg}, we explore the effect of utilizing the CV approach to realistic pre- and post-reconstruction mock catalogs of DESI-like LRGs at $z = 0.8$. We find very good agreement between the predicted and achieved performance of the method as well as excellent agreement between the raw and CV-reduced mean signals.
    \item In Figs.~\ref{fig:xi_elg}, we apply the ZCV and LCV techniques in configuration space to mock ELG samples, finding consistent gains in the signal-to-noise between the two cases. We do not see significant biases given the large error bars and volume limitations of our simulations.
    \item In Section~\ref{sec:bao} and Fig.~\ref{fig:random_draw}, we focus on the application of CV-reduced measurements to BAO analysis. In particular, we comment on the ability of our method to reduce the constraints on the BAO parameters of interest, $\alpha$ and $\epsilon$, and yield a higher-precision measurement that allow us to stress-test the DESI analysis pipeline \citep[see][]{Quintero2023,Fernandez2023}. We find that the noise reduction on the parameter constraints is consistent with the Fisher prediction given the limited number of \textsc{AbacusSummit} simulations (25) we have at hand, with a reduction factor of about 1.5 for $\alpha$ and $\epsilon$.
    \item We release a robust and easy-to-use package for applying CV reduction to mocks generated with the \textsc{AbacusSummit} suite\footnote{\href{https://github.com/abacusorg/abacusutils/}{https://github.com/abacusorg/abacusutils/}}. We also test the performance of our code against the widely used \texttt{nbodykit} package in App.~\ref{app:python}. 
\end{itemize}

We foresee that the CV technique will play an important role in the near future, both as a tool for testing the analysis pipelines of large-scale structure experiments at unprecedented levels of precision,
and as a tool for developing simulation-based theoretical models, which will be of vital importance, as our theoretical errors begin to dominate over systematic ones.

\begin{acknowledgments}
We would like to thank Samuel Brieden, Zhejie Ding, Simone Ferraro, Arnaud de Mattia, Nick Kokron and Chris Blake for illuminating discussions over the course of this project.

Data points for the figures are available at \url{https://doi.org/10.5281/zenodo.8197868}.

This material is based upon work supported by the U.S. Department of Energy (DOE), Office of Science, Office of High Energy Physics of the U.S. Department of Energy under Contract No. DE–AC02–05CH11231, and by the National Energy Research Scientific Computing Center, a DOE Office of Science User Facility under the same contract. Additional support for DESI is provided by the U.S. National Science Foundation, Division of Astronomical Sciences under Contract No. AST-0950945 to the NSF’s National Optical-Infrared Astronomy Research Laboratory; the Science and Technologies Facilities Council of the United Kingdom; the Gordon and Betty Moore Foundation; the Heising-Simons Foundation; the French Alternative Energies and Atomic Energy Commission (CEA); the National Council of Science and Technology of Mexico (CONACYT); the Ministry of Science and Innovation of Spain (MICINN), and by the DESI Member Institutions: \url{https://www.desi.lbl.gov/collaborating-institutions}. Any opinions, findings, and conclusions or recommendations expressed in this material are those of the author(s) and do not necessarily reflect the views of the U. S. National Science Foundation, the U. S. Department of Energy, or any of the listed funding agencies.

The authors are honored to be permitted to conduct scientific research on Iolkam Du’ag (Kitt Peak), a mountain with particular significance to the Tohono O’odham Nation.
\end{acknowledgments}

\bibliographystyle{mnras}
\bibliography{main}

\begin{thebibliography}{}
\makeatletter
\relax
\def\mn@urlcharsother{\let\do\@makeother \do\$\do\&\do\#\do\^\do\_\do\%\do\~}
\def\mn@doi{\begingroup\mn@urlcharsother \@ifnextchar [ {\mn@doi@}
  {\mn@doi@[]}}
\def\mn@doi@[#1]#2{\def\@tempa{#1}\ifx\@tempa\@empty \href
  {http://dx.doi.org/#2} {doi:#2}\else \href {http://dx.doi.org/#2} {#1}\fi
  \endgroup}
\def\mn@eprint#1#2{\mn@eprint@#1:#2::\@nil}
\def\mn@eprint@arXiv#1{\href {http://arxiv.org/abs/#1} {{\tt arXiv:#1}}}
\def\mn@eprint@dblp#1{\href {http://dblp.uni-trier.de/rec/bibtex/#1.xml}
  {dblp:#1}}
\def\mn@eprint@#1:#2:#3:#4\@nil{\def\@tempa {#1}\def\@tempb {#2}\def\@tempc
  {#3}\ifx \@tempc \@empty \let \@tempc \@tempb \let \@tempb \@tempa \fi \ifx
  \@tempb \@empty \def\@tempb {arXiv}\fi \@ifundefined
  {mn@eprint@\@tempb}{\@tempb:\@tempc}{\expandafter \expandafter \csname
  mn@eprint@\@tempb\endcsname \expandafter{\@tempc}}}

\bibitem[\protect\citeauthoryear{{Alam} \& {DESI Collaboration}}{{Alam} \&
  {DESI Collaboration}}{2023}]{Alam2023}
{Alam} S.,  {DESI Collaboration} T.,  2023, in prep.

\bibitem[\protect\citeauthoryear{{Amendola} et~al.,}{{Amendola}
  et~al.}{2013}]{2013LRR....16....6A}
{Amendola} L.,  et~al., 2013, \mn@doi [Living Reviews in Relativity]
  {10.12942/lrr-2013-6}, \href
  {https://ui.adsabs.harvard.edu/abs/2013LRR....16....6A} {16, 6}

\bibitem[\protect\citeauthoryear{{Anderson} et~al.,}{{Anderson}
  et~al.}{2014}]{Anderson14}
{Anderson} L.,  et~al., 2014, \mn@doi [\mnras] {10.1093/mnras/stu523}, \href
  {https://ui.adsabs.harvard.edu/abs/2014MNRAS.441...24A} {441, 24}

\bibitem[\protect\citeauthoryear{{Angulo}, {Zennaro}, {Contreras}, {Aric{\`o}},
  {Pellejero-Iba{\~n}ez}  \& {St{\"u}cker}}{{Angulo}
  et~al.}{2021}]{2021MNRAS.507.5869A}
{Angulo} R.~E.,  {Zennaro} M.,  {Contreras} S.,  {Aric{\`o}} G.,
  {Pellejero-Iba{\~n}ez} M.,   {St{\"u}cker} J.,  2021, \mn@doi [\mnras]
  {10.1093/mnras/stab2018}, \href
  {https://ui.adsabs.harvard.edu/abs/2021MNRAS.507.5869A} {507, 5869}

\bibitem[\protect\citeauthoryear{Beutler et~al.,}{Beutler
  et~al.}{2016}]{Beutler17}
Beutler F.,  et~al., 2016, \mn@doi [Monthly Notices of the Royal Astronomical
  Society] {10.1093/mnras/stw2373}, 464, 3409

\bibitem[\protect\citeauthoryear{{Bouchet}, {Colombi}, {Hivon}  \&
  {Juszkiewicz}}{{Bouchet} et~al.}{1995}]{Bouchet95}
{Bouchet} F.~R.,  {Colombi} S.,  {Hivon} E.,   {Juszkiewicz} R.,  1995, \mn@doi
  [\aap] {10.48550/arXiv.astro-ph/9406013}, \href
  {https://ui.adsabs.harvard.edu/abs/1995A&A...296..575B} {296, 575}

\bibitem[\protect\citeauthoryear{{Buchert}}{{Buchert}}{1989}]{Buchert89}
{Buchert} T.,  1989, \aap, \href
  {https://ui.adsabs.harvard.edu/abs/1989A&A...223....9B} {223, 9}

\bibitem[\protect\citeauthoryear{{Buchert} \& {G{\"o}tz}}{{Buchert} \&
  {G{\"o}tz}}{1987}]{Buchert87}
{Buchert} T.,  {G{\"o}tz} G.,  1987, \mn@doi [Journal of Mathematical Physics]
  {10.1063/1.527717}, \href
  {https://ui.adsabs.harvard.edu/abs/1987JMP....28.2714B} {28, 2714}

\bibitem[\protect\citeauthoryear{{Carlson}, {Reid}  \& {White}}{{Carlson}
  et~al.}{2013}]{Carlson13}
{Carlson} J.,  {Reid} B.,   {White} M.,  2013, \mn@doi [\mnras]
  {10.1093/mnras/sts457}, \href
  {https://ui.adsabs.harvard.edu/abs/2013MNRAS.429.1674C} {429, 1674}

\bibitem[\protect\citeauthoryear{{Carrasco}, {Hertzberg}  \&
  {Senatore}}{{Carrasco} et~al.}{2012}]{2012JHEP...09..082C}
{Carrasco} J. J.~M.,  {Hertzberg} M.~P.,   {Senatore} L.,  2012, \mn@doi
  [Journal of High Energy Physics] {10.1007/JHEP09(2012)082}, \href
  {https://ui.adsabs.harvard.edu/abs/2012JHEP...09..082C} {2012, 82}

\bibitem[\protect\citeauthoryear{{Carretero}, {Castander}, {Gazta{\~n}aga},
  {Crocce}  \& {Fosalba}}{{Carretero} et~al.}{2015}]{2015MNRAS.447..646C}
{Carretero} J.,  {Castander} F.~J.,  {Gazta{\~n}aga} E.,  {Crocce} M.,
  {Fosalba} P.,  2015, \mn@doi [\mnras] {10.1093/mnras/stu2402}, \href
  {https://ui.adsabs.harvard.edu/abs/2015MNRAS.447..646C} {447, 646}

\bibitem[\protect\citeauthoryear{{Chartier} \& {Wandelt}}{{Chartier} \&
  {Wandelt}}{2022}]{2022MNRAS.509.2220C}
{Chartier} N.,  {Wandelt} B.~D.,  2022, \mn@doi [\mnras]
  {10.1093/mnras/stab3097}, \href
  {https://ui.adsabs.harvard.edu/abs/2022MNRAS.509.2220C} {509, 2220}

\bibitem[\protect\citeauthoryear{{Chartier}, {Wandelt}, {Akrami}  \&
  {Villaescusa-Navarro}}{{Chartier} et~al.}{2021}]{2021MNRAS.503.1897C}
{Chartier} N.,  {Wandelt} B.,  {Akrami} Y.,   {Villaescusa-Navarro} F.,  2021,
  \mn@doi [\mnras] {10.1093/mnras/stab430}, \href
  {https://ui.adsabs.harvard.edu/abs/2021MNRAS.503.1897C} {503, 1897}

\bibitem[\protect\citeauthoryear{{Chen} \& {DESI Collaboration}}{{Chen} \&
  {DESI Collaboration}}{2023}]{Chen2023}
{Chen} S.-F.,  {DESI Collaboration} T.,  2023, in prep.

\bibitem[\protect\citeauthoryear{{Chen}, {Vlah}  \& {White}}{{Chen}
  et~al.}{2019}]{2019JCAP...09..017C}
{Chen} S.-F.,  {Vlah} Z.,   {White} M.,  2019, \mn@doi [\jcap]
  {10.1088/1475-7516/2019/09/017}, \href
  {https://ui.adsabs.harvard.edu/abs/2019JCAP...09..017C} {2019, 017}

\bibitem[\protect\citeauthoryear{{Chen}, {Vlah}, {Castorina}  \&
  {White}}{{Chen} et~al.}{2021}]{Chen21}
{Chen} S.-F.,  {Vlah} Z.,  {Castorina} E.,   {White} M.,  2021, \mn@doi [\jcap]
  {10.1088/1475-7516/2021/03/100}, \href
  {https://ui.adsabs.harvard.edu/abs/2021JCAP...03..100C} {2021, 100}

\bibitem[\protect\citeauthoryear{{Chen}, {Vlah}  \& {White}}{{Chen}
  et~al.}{2022}]{Chen22}
{Chen} S.-F.,  {Vlah} Z.,   {White} M.,  2022, \mn@doi [\jcap]
  {10.1088/1475-7516/2022/02/008}, \href
  {https://ui.adsabs.harvard.edu/abs/2022JCAP...02..008C} {2022, 008}

\bibitem[\protect\citeauthoryear{{Chen}, {Ding}, {Paillas}  \& {DESI
  Collaboration}}{{Chen} et~al.}{2023}]{ChenDingPaillas2023}
{Chen} X.,  {Ding} Z.,  {Paillas} E.,   {DESI Collaboration} T.,  2023, in
  prep.

\bibitem[\protect\citeauthoryear{{Chisari} \& {Pontzen}}{{Chisari} \&
  {Pontzen}}{2019}]{2019PhRvD.100b3543C}
{Chisari} N.~E.,  {Pontzen} A.,  2019, \mn@doi [\prd]
  {10.1103/PhysRevD.100.023543}, \href
  {https://ui.adsabs.harvard.edu/abs/2019PhRvD.100b3543C} {100, 023543}

\bibitem[\protect\citeauthoryear{{Chuang}, {Kitaura}, {Prada}, {Zhao}  \&
  {Yepes}}{{Chuang} et~al.}{2015}]{2015MNRAS.446.2621C}
{Chuang} C.-H.,  {Kitaura} F.-S.,  {Prada} F.,  {Zhao} C.,   {Yepes} G.,  2015,
  \mn@doi [\mnras] {10.1093/mnras/stu2301}, \href
  {https://ui.adsabs.harvard.edu/abs/2015MNRAS.446.2621C} {446, 2621}

\bibitem[\protect\citeauthoryear{{Cohn}, {White}, {Chang}, {Holder},
  {Padmanabhan}  \& {Dor{\'e}}}{{Cohn} et~al.}{2016}]{2016MNRAS.457.2068C}
{Cohn} J.~D.,  {White} M.,  {Chang} T.-C.,  {Holder} G.,  {Padmanabhan} N.,
  {Dor{\'e}} O.,  2016, \mn@doi [\mnras] {10.1093/mnras/stw108}, \href
  {https://ui.adsabs.harvard.edu/abs/2016MNRAS.457.2068C} {457, 2068}

\bibitem[\protect\citeauthoryear{{Coles}, {Melott}  \& {Shandarin}}{{Coles}
  et~al.}{1993}]{Coles93}
{Coles} P.,  {Melott} A.~L.,   {Shandarin} S.~F.,  1993, \mn@doi [\mnras]
  {10.1093/mnras/260.4.765}, \href
  {https://ui.adsabs.harvard.edu/abs/1993MNRAS.260..765C} {260, 765}

\bibitem[\protect\citeauthoryear{{Crocce} \& {Scoccimarro}}{{Crocce} \&
  {Scoccimarro}}{2006}]{2006PhRvD..73f3519C}
{Crocce} M.,  {Scoccimarro} R.,  2006, \mn@doi [\prd]
  {10.1103/PhysRevD.73.063519}, \href
  {https://ui.adsabs.harvard.edu/abs/2006PhRvD..73f3519C} {73, 063519}

\bibitem[\protect\citeauthoryear{{Crocce}, {Castander}, {Gazta{\~n}aga},
  {Fosalba}  \& {Carretero}}{{Crocce} et~al.}{2015}]{2015MNRAS.453.1513C}
{Crocce} M.,  {Castander} F.~J.,  {Gazta{\~n}aga} E.,  {Fosalba} P.,
  {Carretero} J.,  2015, \mn@doi [\mnras] {10.1093/mnras/stv1708}, \href
  {https://ui.adsabs.harvard.edu/abs/2015MNRAS.453.1513C} {453, 1513}

\bibitem[\protect\citeauthoryear{{DESI Collaboration} et~al.,}{{DESI
  Collaboration} et~al.}{2016a}]{2016arXiv161100036D}
{DESI Collaboration} et~al., 2016a, \mn@doi [arXiv e-prints]
  {10.48550/arXiv.1611.00036}, \href
  {https://ui.adsabs.harvard.edu/abs/2016arXiv161100036D} {p. arXiv:1611.00036}

\bibitem[\protect\citeauthoryear{{DESI Collaboration} et~al.,}{{DESI
  Collaboration} et~al.}{2016b}]{2016arXiv161100037D}
{DESI Collaboration} et~al., 2016b, \mn@doi [arXiv e-prints]
  {10.48550/arXiv.1611.00037}, \href
  {https://ui.adsabs.harvard.edu/abs/2016arXiv161100037D} {p. arXiv:1611.00037}

\bibitem[\protect\citeauthoryear{{DESI Collaboration} et~al.,}{{DESI
  Collaboration} et~al.}{2022}]{2022AJ....164..207D}
{DESI Collaboration} et~al., 2022, \mn@doi [\aj] {10.3847/1538-3881/ac882b},
  \href {https://ui.adsabs.harvard.edu/abs/2022AJ....164..207D} {164, 207}

\bibitem[\protect\citeauthoryear{{DeRose} et~al.,}{{DeRose}
  et~al.}{2019}]{2019ApJ...875...69D}
{DeRose} J.,  et~al., 2019, \mn@doi [\apj] {10.3847/1538-4357/ab1085}, \href
  {https://ui.adsabs.harvard.edu/abs/2019ApJ...875...69D} {875, 69}

\bibitem[\protect\citeauthoryear{{DeRose} et~al.,}{{DeRose}
  et~al.}{2023a}]{2023arXiv230309762D}
{DeRose} J.,  et~al., 2023a, \mn@doi [arXiv e-prints]
  {10.48550/arXiv.2303.09762}, \href
  {https://ui.adsabs.harvard.edu/abs/2023arXiv230309762D} {p. arXiv:2303.09762}

\bibitem[\protect\citeauthoryear{{DeRose}, {Chen}, {Kokron}  \&
  {White}}{{DeRose} et~al.}{2023b}]{DeRose23}
{DeRose} J.,  {Chen} S.-F.,  {Kokron} N.,   {White} M.,  2023b, \mn@doi [\jcap]
  {10.1088/1475-7516/2023/02/008}, \href
  {https://ui.adsabs.harvard.edu/abs/2023JCAP...02..008D} {2023, 008}

\bibitem[\protect\citeauthoryear{{Ding}, {Seo}, {Vlah}, {Feng}, {Schmittfull}
  \& {Beutler}}{{Ding} et~al.}{2018}]{2018MNRAS.479.1021D}
{Ding} Z.,  {Seo} H.-J.,  {Vlah} Z.,  {Feng} Y.,  {Schmittfull} M.,   {Beutler}
  F.,  2018, \mn@doi [\mnras] {10.1093/mnras/sty1413}, \href
  {https://ui.adsabs.harvard.edu/abs/2018MNRAS.479.1021D} {479, 1021}

\bibitem[\protect\citeauthoryear{{Ding} et~al.,}{{Ding}
  et~al.}{2022}]{2022MNRAS.514.3308D}
{Ding} Z.,  et~al., 2022, \mn@doi [\mnras] {10.1093/mnras/stac1501}, \href
  {https://ui.adsabs.harvard.edu/abs/2022MNRAS.514.3308D} {514, 3308}

\bibitem[\protect\citeauthoryear{{Dodelson} \& {Schmidt}}{{Dodelson} \&
  {Schmidt}}{2020}]{Dodelson20}
{Dodelson} S.,  {Schmidt} F.,  2020, {Modern Cosmology},
  \mn@doi{10.1016/C2017-0-01943-2.
}, \url {https://doi.org/10.1016/C2017-0-01943-2}

\bibitem[\protect\citeauthoryear{{Dor{\'e}} et~al.,}{{Dor{\'e}}
  et~al.}{2014}]{2014arXiv1412.4872D}
{Dor{\'e}} O.,  et~al., 2014, \mn@doi [arXiv e-prints]
  {10.48550/arXiv.1412.4872}, \href
  {https://ui.adsabs.harvard.edu/abs/2014arXiv1412.4872D} {p. arXiv:1412.4872}

\bibitem[\protect\citeauthoryear{{Dor{\'e}} et~al.,}{{Dor{\'e}}
  et~al.}{2018}]{2018arXiv180505489D}
{Dor{\'e}} O.,  et~al., 2018, \mn@doi [arXiv e-prints]
  {10.48550/arXiv.1805.05489}, \href
  {https://ui.adsabs.harvard.edu/abs/2018arXiv180505489D} {p. arXiv:1805.05489}

\bibitem[\protect\citeauthoryear{{Doroshkevich}, {Zeldovich}, {Syunyaev}  \&
  {Khlopov}}{{Doroshkevich} et~al.}{1980}]{Doroshkevich80}
{Doroshkevich} A.~G.,  {Zeldovich} Y.~B.,  {Syunyaev} R.~A.,   {Khlopov} M.~Y.,
   1980, Pisma v Astronomicheskii Zhurnal, \href
  {https://ui.adsabs.harvard.edu/abs/1980PAZh....6..457D} {6, 457}

\bibitem[\protect\citeauthoryear{{Eisenstein}, {Seo}, {Sirko}  \&
  {Spergel}}{{Eisenstein} et~al.}{2007a}]{ESSS07}
{Eisenstein} D.~J.,  {Seo} H.-J.,  {Sirko} E.,   {Spergel} D.~N.,  2007a,
  \mn@doi [\apj] {10.1086/518712}, \href
  {http://adsabs.harvard.edu/abs/2007ApJ...664..675E} {664, 675}

\bibitem[\protect\citeauthoryear{{Eisenstein}, {Seo}  \& {White}}{{Eisenstein}
  et~al.}{2007b}]{ESW07}
{Eisenstein} D.~J.,  {Seo} H.-J.,   {White} M.,  2007b, \mn@doi [\apj]
  {10.1086/518755}, \href
  {https://ui.adsabs.harvard.edu/abs/2007ApJ...664..660E} {664, 660}

\bibitem[\protect\citeauthoryear{{Euclid Collaboration} et~al.,}{{Euclid
  Collaboration} et~al.}{2019}]{2019MNRAS.484.5509E}
{Euclid Collaboration} et~al., 2019, \mn@doi [\mnras] {10.1093/mnras/stz197},
  \href {https://ui.adsabs.harvard.edu/abs/2019MNRAS.484.5509E} {484, 5509}

\bibitem[\protect\citeauthoryear{{Euclid Collaboration} et~al.,}{{Euclid
  Collaboration} et~al.}{2022}]{2022A&A...662A.112E}
{Euclid Collaboration} et~al., 2022, \mn@doi [\aap]
  {10.1051/0004-6361/202141938}, \href
  {https://ui.adsabs.harvard.edu/abs/2022A&A...662A.112E} {662, A112}

\bibitem[\protect\citeauthoryear{{Feng}, {Chu}, {Seljak}  \& {McDonald}}{{Feng}
  et~al.}{2016}]{2016MNRAS.463.2273F}
{Feng} Y.,  {Chu} M.-Y.,  {Seljak} U.,   {McDonald} P.,  2016, \mn@doi [\mnras]
  {10.1093/mnras/stw2123}, \href
  {https://ui.adsabs.harvard.edu/abs/2016MNRAS.463.2273F} {463, 2273}

\bibitem[\protect\citeauthoryear{{Fern\'andez} \& {DESI
  Collaboration}}{{Fern\'andez} \& {DESI Collaboration}}{2023}]{Fernandez2023}
{Fern\'andez} J.~M.,  {DESI Collaboration} T.,  2023, in prep.

\bibitem[\protect\citeauthoryear{{Fosalba}, {Gazta{\~n}aga}, {Castander}  \&
  {Crocce}}{{Fosalba} et~al.}{2015a}]{2015MNRAS.447.1319F}
{Fosalba} P.,  {Gazta{\~n}aga} E.,  {Castander} F.~J.,   {Crocce} M.,  2015a,
  \mn@doi [\mnras] {10.1093/mnras/stu2464}, \href
  {https://ui.adsabs.harvard.edu/abs/2015MNRAS.447.1319F} {447, 1319}

\bibitem[\protect\citeauthoryear{{Fosalba}, {Crocce}, {Gazta{\~n}aga}  \&
  {Castander}}{{Fosalba} et~al.}{2015b}]{2015MNRAS.448.2987F}
{Fosalba} P.,  {Crocce} M.,  {Gazta{\~n}aga} E.,   {Castander} F.~J.,  2015b,
  \mn@doi [\mnras] {10.1093/mnras/stv138}, \href
  {https://ui.adsabs.harvard.edu/abs/2015MNRAS.448.2987F} {448, 2987}

\bibitem[\protect\citeauthoryear{{Garcia-Quintero} \& {DESI
  Collaboration}}{{Garcia-Quintero} \& {DESI
  Collaboration}}{2023}]{Quintero2023}
{Garcia-Quintero} C.,  {DESI Collaboration} T.,  2023, in prep.

\bibitem[\protect\citeauthoryear{{Garrison}, {Eisenstein}, {Ferrer}, {Tinker},
  {Pinto}  \& {Weinberg}}{{Garrison} et~al.}{2018}]{2018ApJS..236...43G}
{Garrison} L.~H.,  {Eisenstein} D.~J.,  {Ferrer} D.,  {Tinker} J.~L.,  {Pinto}
  P.~A.,   {Weinberg} D.~H.,  2018, \mn@doi [\apjs] {10.3847/1538-4365/aabfd3},
  \href {https://ui.adsabs.harvard.edu/abs/2018ApJS..236...43G} {236, 43}

\bibitem[\protect\citeauthoryear{{Garrison}, {Eisenstein}  \&
  {Pinto}}{{Garrison} et~al.}{2019}]{2019MNRAS.485.3370G}
{Garrison} L.~H.,  {Eisenstein} D.~J.,   {Pinto} P.~A.,  2019, \mn@doi [\mnras]
  {10.1093/mnras/stz634}, \href
  {https://ui.adsabs.harvard.edu/abs/2019MNRAS.485.3370G} {485, 3370}

\bibitem[\protect\citeauthoryear{{Garrison}, {Eisenstein}, {Ferrer},
  {Maksimova}  \& {Pinto}}{{Garrison} et~al.}{2021}]{2021MNRAS.508..575G}
{Garrison} L.~H.,  {Eisenstein} D.~J.,  {Ferrer} D.,  {Maksimova} N.~A.,
  {Pinto} P.~A.,  2021, \mn@doi [\mnras] {10.1093/mnras/stab2482}, \href
  {https://ui.adsabs.harvard.edu/abs/2021MNRAS.508..575G} {508, 575}

\bibitem[\protect\citeauthoryear{{Goroff}, {Grinstein}, {Rey}  \&
  {Wise}}{{Goroff} et~al.}{1986}]{1986ApJ...311....6G}
{Goroff} M.~H.,  {Grinstein} B.,  {Rey} S.~J.,   {Wise} M.~B.,  1986, \mn@doi
  [\apj] {10.1086/164749}, \href
  {https://ui.adsabs.harvard.edu/abs/1986ApJ...311....6G} {311, 6}

\bibitem[\protect\citeauthoryear{{Hadzhiyska}, {Eisenstein}, {Bose}, {Garrison}
   \& {Maksimova}}{{Hadzhiyska} et~al.}{2022}]{2022MNRAS.509..501H}
{Hadzhiyska} B.,  {Eisenstein} D.,  {Bose} S.,  {Garrison} L.~H.,   {Maksimova}
  N.,  2022, \mn@doi [\mnras] {10.1093/mnras/stab2980}, \href
  {https://ui.adsabs.harvard.edu/abs/2022MNRAS.509..501H} {509, 501}

\bibitem[\protect\citeauthoryear{{Hand}, {Feng}, {Beutler}, {Li}, {Modi},
  {Seljak}  \& {Slepian}}{{Hand} et~al.}{2018}]{2018AJ....156..160H}
{Hand} N.,  {Feng} Y.,  {Beutler} F.,  {Li} Y.,  {Modi} C.,  {Seljak} U.,
  {Slepian} Z.,  2018, \mn@doi [\aj] {10.3847/1538-3881/aadae0}, \href
  {https://ui.adsabs.harvard.edu/abs/2018AJ....156..160H} {156, 160}

\bibitem[\protect\citeauthoryear{{Heitmann} et~al.,}{{Heitmann}
  et~al.}{2019}]{2019ApJS..245...16H}
{Heitmann} K.,  et~al., 2019, \mn@doi [\apjs] {10.3847/1538-4365/ab4da1}, \href
  {https://ui.adsabs.harvard.edu/abs/2019ApJS..245...16H} {245, 16}

\bibitem[\protect\citeauthoryear{{Hockney} \& {Eastwood}}{{Hockney} \&
  {Eastwood}}{1981}]{1981csup.book.....H}
{Hockney} R.~W.,  {Eastwood} J.~W.,  1981, {Computer Simulation Using
  Particles}

\bibitem[\protect\citeauthoryear{{Ivanov}}{{Ivanov}}{2022}]{Ivanov22}
{Ivanov} M.~M.,  2022, \mn@doi [arXiv e-prints] {10.48550/arXiv.2212.08488},
  \href {https://ui.adsabs.harvard.edu/abs/2022arXiv221208488I} {p.
  arXiv:2212.08488}

\bibitem[\protect\citeauthoryear{{Ivezi{\'c}} et~al.,}{{Ivezi{\'c}}
  et~al.}{2019}]{2019ApJ...873..111I}
{Ivezi{\'c}} {\v{Z}}.,  et~al., 2019, \mn@doi [\apj]
  {10.3847/1538-4357/ab042c}, \href
  {https://ui.adsabs.harvard.edu/abs/2019ApJ...873..111I} {873, 111}

\bibitem[\protect\citeauthoryear{{Jain} \& {Bertschinger}}{{Jain} \&
  {Bertschinger}}{1994}]{1994ApJ...431..495J}
{Jain} B.,  {Bertschinger} E.,  1994, \mn@doi [\apj] {10.1086/174502}, \href
  {https://ui.adsabs.harvard.edu/abs/1994ApJ...431..495J} {431, 495}

\bibitem[\protect\citeauthoryear{{Kokron}, {Chen}, {White}, {DeRose}  \&
  {Maus}}{{Kokron} et~al.}{2022}]{Kokron22b}
{Kokron} N.,  {Chen} S.-F.,  {White} M.,  {DeRose} J.,   {Maus} M.,  2022,
  \mn@doi [\jcap] {10.1088/1475-7516/2022/09/059}, \href
  {https://ui.adsabs.harvard.edu/abs/2022JCAP...09..059K} {2022, 059}

\bibitem[\protect\citeauthoryear{{LSST Dark Energy Science
  Collaboration}}{{LSST Dark Energy Science
  Collaboration}}{2012}]{2012arXiv1211.0310L}
{LSST Dark Energy Science Collaboration} 2012, \mn@doi [arXiv e-prints]
  {10.48550/arXiv.1211.0310}, \href
  {https://ui.adsabs.harvard.edu/abs/2012arXiv1211.0310L} {p. arXiv:1211.0310}

\bibitem[\protect\citeauthoryear{Lam, Pitrou  \& Seibert}{Lam
  et~al.}{2015}]{10.1145/2833157.2833162}
Lam S.~K.,  Pitrou A.,   Seibert S.,  2015, in Proceedings of the Second
  Workshop on the LLVM Compiler Infrastructure in HPC. LLVM '15.
Association for Computing Machinery, New York, NY, USA,
  \mn@doi{10.1145/2833157.2833162}, \url
  {https://doi.org/10.1145/2833157.2833162}

\bibitem[\protect\citeauthoryear{{Laureijs} et~al.,}{{Laureijs}
  et~al.}{2011}]{2011arXiv1110.3193L}
{Laureijs} R.,  et~al., 2011, \mn@doi [arXiv e-prints]
  {10.48550/arXiv.1110.3193}, \href
  {https://ui.adsabs.harvard.edu/abs/2011arXiv1110.3193L} {p. arXiv:1110.3193}

\bibitem[\protect\citeauthoryear{{Lesgourgues}}{{Lesgourgues}}{2011}]{2011arXiv1104.2932L}
{Lesgourgues} J.,  2011, \mn@doi [arXiv e-prints] {10.48550/arXiv.1104.2932},
  \href {https://ui.adsabs.harvard.edu/abs/2011arXiv1104.2932L} {p.
  arXiv:1104.2932}

\bibitem[\protect\citeauthoryear{{Levi} et~al.,}{{Levi}
  et~al.}{2013}]{2013arXiv1308.0847L}
{Levi} M.,  et~al., 2013, \mn@doi [arXiv e-prints] {10.48550/arXiv.1308.0847},
  \href {https://ui.adsabs.harvard.edu/abs/2013arXiv1308.0847L} {p.
  arXiv:1308.0847}

\bibitem[\protect\citeauthoryear{{Levi} et~al.,}{{Levi}
  et~al.}{2019}]{2019BAAS...51g..57L}
{Levi} M.,  et~al., 2019, in Bulletin of the American Astronomical Society.
  p.~57 (\mn@eprint {arXiv} {1907.10688}), \mn@doi{10.48550/arXiv.1907.10688}

\bibitem[\protect\citeauthoryear{{Maksimova}, {Garrison}, {Eisenstein},
  {Hadzhiyska}, {Bose}  \& {Satterthwaite}}{{Maksimova}
  et~al.}{2021}]{2021MNRAS.508.4017M}
{Maksimova} N.~A.,  {Garrison} L.~H.,  {Eisenstein} D.~J.,  {Hadzhiyska} B.,
  {Bose} S.,   {Satterthwaite} T.~P.,  2021, \mn@doi [\mnras]
  {10.1093/mnras/stab2484}, \href
  {https://ui.adsabs.harvard.edu/abs/2021MNRAS.508.4017M} {508, 4017}

\bibitem[\protect\citeauthoryear{{Matsubara}}{{Matsubara}}{2008}]{Matsubara08}
{Matsubara} T.,  2008, \mn@doi [\prd] {10.1103/PhysRevD.77.063530}, \href
  {https://ui.adsabs.harvard.edu/abs/2008PhRvD..77f3530M} {77, 063530}

\bibitem[\protect\citeauthoryear{Owen}{Owen}{2013}]{mcbook}
Owen A.~B.,  2013, Monte Carlo theory, methods and examples

\bibitem[\protect\citeauthoryear{Padmanabhan \& White}{Padmanabhan \&
  White}{2008}]{Padmanabhan08}
Padmanabhan N.,  White M.,  2008, \mn@doi [Phys. Rev. D]
  {10.1103/PhysRevD.77.123540}, 77, 123540

\bibitem[\protect\citeauthoryear{{Padmanabhan}, {Xu}, {Eisenstein}, {Scalzo},
  {Cuesta}, {Mehta}  \& {Kazin}}{{Padmanabhan} et~al.}{2012}]{Padmanabhan12}
{Padmanabhan} N.,  {Xu} X.,  {Eisenstein} D.~J.,  {Scalzo} R.,  {Cuesta} A.~J.,
   {Mehta} K.~T.,   {Kazin} E.,  2012, \mn@doi [\mnras]
  {10.1111/j.1365-2966.2012.21888.x}, \href
  {https://ui.adsabs.harvard.edu/abs/2012MNRAS.427.2132P} {427, 2132}

\bibitem[\protect\citeauthoryear{{Pauls} \& {Melott}}{{Pauls} \&
  {Melott}}{1995}]{Pauls95}
{Pauls} J.~L.,  {Melott} A.~L.,  1995, \mn@doi [\mnras]
  {10.1093/mnras/274.1.99}, \href
  {https://ui.adsabs.harvard.edu/abs/1995MNRAS.274...99P} {274, 99}

\bibitem[\protect\citeauthoryear{{Perko}, {Senatore}, {Jennings}  \&
  {Wechsler}}{{Perko} et~al.}{2016}]{2016arXiv161009321P}
{Perko} A.,  {Senatore} L.,  {Jennings} E.,   {Wechsler} R.~H.,  2016, \mn@doi
  [arXiv e-prints] {10.48550/arXiv.1610.09321}, \href
  {https://ui.adsabs.harvard.edu/abs/2016arXiv161009321P} {p. arXiv:1610.09321}

\bibitem[\protect\citeauthoryear{{Pieri} et~al.,}{{Pieri}
  et~al.}{2016}]{2016sf2a.conf..259P}
{Pieri} M.~M.,  et~al., 2016, in {Reyl{\'e}} C.,  {Richard} J.,  {Cambr{\'e}sy}
  L.,  {Deleuil} M.,  {P{\'e}contal} E.,  {Tresse} L.,   {Vauglin} I.,  eds,
  SF2A-2016: Proceedings of the Annual meeting of the French Society of
  Astronomy and Astrophysics. pp 259--266 (\mn@eprint {arXiv} {1611.09388}),
  \mn@doi{10.48550/arXiv.1611.09388}

\bibitem[\protect\citeauthoryear{{Seo}, {Beutler}, {Ross}  \& {Saito}}{{Seo}
  et~al.}{2016}]{2016MNRAS.460.2453S}
{Seo} H.-J.,  {Beutler} F.,  {Ross} A.~J.,   {Saito} S.,  2016, \mn@doi
  [\mnras] {10.1093/mnras/stw1138}, \href
  {https://ui.adsabs.harvard.edu/abs/2016MNRAS.460.2453S} {460, 2453}

\bibitem[\protect\citeauthoryear{{Silber} et~al.,}{{Silber}
  et~al.}{2023}]{2023AJ....165....9S}
{Silber} J.~H.,  et~al., 2023, \mn@doi [\aj] {10.3847/1538-3881/ac9ab1}, \href
  {https://ui.adsabs.harvard.edu/abs/2023AJ....165....9S} {165, 9}

\bibitem[\protect\citeauthoryear{{Spergel} et~al.,}{{Spergel}
  et~al.}{2015}]{2015arXiv150303757S}
{Spergel} D.,  et~al., 2015, \mn@doi [arXiv e-prints]
  {10.48550/arXiv.1503.03757}, \href
  {https://ui.adsabs.harvard.edu/abs/2015arXiv150303757S} {p. arXiv:1503.03757}

\bibitem[\protect\citeauthoryear{{Tamura} et~al.,}{{Tamura}
  et~al.}{2016}]{2016SPIE.9908E..1MT}
{Tamura} N.,  et~al., 2016, in {Evans} C.~J.,  {Simard} L.,   {Takami} H.,
  eds,  Society of Photo-Optical Instrumentation Engineers (SPIE) Conference
  Series Vol. 9908, Ground-based and Airborne Instrumentation for Astronomy VI.
  p. 99081M (\mn@eprint {arXiv} {1608.01075}), \mn@doi{10.1117/12.2232103}

\bibitem[\protect\citeauthoryear{{Tassev}}{{Tassev}}{2014}]{Tassev14}
{Tassev} S.,  2014, \mn@doi [\jcap] {10.1088/1475-7516/2014/06/008}, \href
  {https://ui.adsabs.harvard.edu/abs/2014JCAP...06..008T} {2014, 008}

\bibitem[\protect\citeauthoryear{{To} et~al.,}{{To}
  et~al.}{2023}]{2023arXiv230312104T}
{To} C.-H.,  et~al., 2023, \mn@doi [arXiv e-prints]
  {10.48550/arXiv.2303.12104}, \href
  {https://ui.adsabs.harvard.edu/abs/2023arXiv230312104T} {p. arXiv:2303.12104}

\bibitem[\protect\citeauthoryear{{Variu} \& {DESI Collaboration}}{{Variu} \&
  {DESI Collaboration}}{2023}]{Variu2023}
{Variu} A.,  {DESI Collaboration} T.,  2023, in prep.

\bibitem[\protect\citeauthoryear{Virtanen et~al.,}{Virtanen
  et~al.}{2020}]{2020SciPy-NMeth}
Virtanen P.,  et~al., 2020, \mn@doi [Nature Methods]
  {10.1038/s41592-019-0686-2}, \href {https://rdcu.be/b08Wh} {17, 261}

\bibitem[\protect\citeauthoryear{{Vlah}, {White}  \& {Aviles}}{{Vlah}
  et~al.}{2015}]{2015JCAP...09..014V}
{Vlah} Z.,  {White} M.,   {Aviles} A.,  2015, \mn@doi [\jcap]
  {10.1088/1475-7516/2015/09/014}, \href
  {https://ui.adsabs.harvard.edu/abs/2015JCAP...09..014V} {2015, 014}

\bibitem[\protect\citeauthoryear{{Wadekar} \& {Scoccimarro}}{{Wadekar} \&
  {Scoccimarro}}{2020}]{2020PhRvD.102l3517W}
{Wadekar} D.,  {Scoccimarro} R.,  2020, \mn@doi [\prd]
  {10.1103/PhysRevD.102.123517}, \href
  {https://ui.adsabs.harvard.edu/abs/2020PhRvD.102l3517W} {102, 123517}

\bibitem[\protect\citeauthoryear{{White}}{{White}}{2014}]{White14}
{White} M.,  2014, \mn@doi [\mnras] {10.1093/mnras/stu209}, \href
  {https://ui.adsabs.harvard.edu/abs/2014MNRAS.439.3630W} {439, 3630}

\bibitem[\protect\citeauthoryear{{White}}{{White}}{2015}]{2015MNRAS.450.3822W}
{White} M.,  2015, \mn@doi [\mnras] {10.1093/mnras/stv842}, \href
  {https://ui.adsabs.harvard.edu/abs/2015MNRAS.450.3822W} {450, 3822}

\bibitem[\protect\citeauthoryear{{Yoshisato}, {Morikawa}, {Gouda}  \&
  {Mouri}}{{Yoshisato} et~al.}{2006}]{Yoshisato06}
{Yoshisato} A.,  {Morikawa} M.,  {Gouda} N.,   {Mouri} H.,  2006, \mn@doi
  [\apj] {10.1086/498496}, \href
  {https://ui.adsabs.harvard.edu/abs/2006ApJ...637..555Y} {637, 555}

\bibitem[\protect\citeauthoryear{{Yuan}, {Garrison}, {Hadzhiyska}, {Bose}  \&
  {Eisenstein}}{{Yuan} et~al.}{2022}]{2022MNRAS.510.3301Y}
{Yuan} S.,  {Garrison} L.~H.,  {Hadzhiyska} B.,  {Bose} S.,   {Eisenstein}
  D.~J.,  2022, \mn@doi [\mnras] {10.1093/mnras/stab3355}, \href
  {https://ui.adsabs.harvard.edu/abs/2022MNRAS.510.3301Y} {510, 3301}

\bibitem[\protect\citeauthoryear{{Zel'dovich}}{{Zel'dovich}}{1970}]{Zeldovich70}
{Zel'dovich} Y.~B.,  1970, \aap, \href
  {https://ui.adsabs.harvard.edu/abs/1970A&A.....5...84Z} {5, 84}

\makeatother
\end{thebibliography}

\appendix

\section{Python package}
\label{app:python}

We implement the core numerical algorithms (HOD generation, power spectrum computation, etc.) used in this work in the open-source \texttt{abacusutils} Python package. Documentation, including tutorials, is available online\footnote{\url{https://abacusutils.readthedocs.io}}. Special attention is paid to parallelization and optimization; we give some of these details for the power spectrum module in this appendix. See \cite{2022MNRAS.510.3301Y} for discussion of the HOD optimization.

\subsection{Triangle-shaped cloud}
To measure the power spectrum of a collection of points, we first assign their mass to a cubic mesh using triangle-shaped cloud (TSC) mass assignment \citep{1981csup.book.....H}.  TSC is computationally expensive because each particle updates a cloud of 27 mesh cells around it which are mostly not contiguous (or even proximate) in memory.  The performance bottleneck is thus the memory bandwidth for random writes.  Typically one CPU cannot saturate the memory bandwidth, especially on multi-socket systems, so parallelization is important. We target shared-memory (i.e.~single node) parallelization, for reasons described below.

The most naive shared-memory parallelization scheme would be to parallelize over particles, with different threads treating different particles, and all threads writing to a shared grid. This is not thread safe, however: if two threads try to update overlapping clouds, a race condition will result.

Instead, the algorithm we implement is to partition the particles along the $x$-dimension into stripes at least as wide as the TSC cloud.  Then all even-parity stripes can be written in parallel, followed by all odd-parity stripes, without any race conditions.  We implement the partitioning as a parallel radix sort which is typically $5$--$10\times$ times faster than the TSC step.

The sort and TSC are implemented in \texttt{numba}\footnote{\url{https://numba.pydata.org/}} \citep{10.1145/2833157.2833162}, a just-in-time compiler for Python and NumPy. \texttt{numba} also enables loop-level parallelization, which is key for our TSC implementation.  We take care to check the data types in the LLVM intermediate representation and modify the Python code as necessary to avoid expensive type promotions.

The thread scaling of this implementation is quite good: for a $256^3$ grid with $10^7$ randomly placed particles ($\bar n = 10^{-3} h^3\ \mathrm{Mpc}^{-3}$ in a $2 \gpcoh$ box) in 32-bit precision, 1 thread takes 930 ms (10 M part/s) and 32 threads take 20 ms (500 M part/s) on a 2 $\times$ 32-core Intel Ice Lake CPU platform (the scaling begins to soften at higher thread count).  This is a super linear scaling which comes from the partitioning---the partitioning for parallelization has the side effect of producing a more efficient particle order.  The serial runtime could thus be improved, but in general the best partitioning will depend on sortedness of the particles, the grid size, and the CPU cache architecture.  The number of partitions is a performance tuning parameter, up to a maximum set by the TSC cloud size.

The TSC code is tested against \texttt{nbodykit} \citep{2018AJ....156..160H}, and it gives the same results. In our testing, \texttt{nbodykit} is slower on the same hardware, though: the single-core performance is 2.1 M part/sec, and the 32-core performance is 30 M part/s, or 16$\times$ slower than \texttt{abacusutils}. On the other hand, \texttt{nbodykit}, as an MPI-backed code, can use multiple nodes and scale to larger problem sizes.

Our implementation targets shared-memory parallelism for several reasons: (1) the problem sizes we are interested in typically fit within a single node's memory; (2) it lets us skip expensive operations like co-adding density grids from different processes, and zeroing said grids; (3) the programming tools for thread-level parallelization are easier to use from a developer perspective and easier to install from a user perspective. Finally, a fast single-node implementation is still useful as part of a larger multi-node implementation, if the problem sizes ever grow to the point where that is necessary.

\subsection{Power spectrum}
The power spectrum implementation uses the SciPy \citep{2020SciPy-NMeth} fast Fourier transform (FFT). While past iterations of the Abacus power spectrum code used FFTW, the SciPy FFT is now as least as fast for the problems sizes in which we are interested.

Our implementation again uses \texttt{numba} for manipulations of the FFT mesh. \texttt{numba}'s procedural (loop-oriented) programming model enables more efficient manipulation of a large mesh, compared with NumPy's array-oriented programming model. The NumPy model often forces the programmer to construct large intermediate arrays (e.g.~a grid of $k$-magnitudes), which consume large amounts of memory and are slow to manipulate. Using \texttt{numba}, an operation like counting the number of modes in 2000 $(k,\mu)$ bins takes 32 ms for a $1024^3$ grid using 64 threads, with excellent thread scaling.

End-to-end computation of a 1D power spectrum of $10^7$ points on a $256^3$ mesh with 100 $k$-bins takes 1.1 seconds (9.2 M part/s) with 1 thread, and 31 ms (320 M part/s) with 32 threads. Using \texttt{nbodykit} takes 5.2 seconds (1.9 M part/s) with 1 thread, and 300 ms (33 M part/s) with 32 threads, or about $10\times$ slower than \texttt{abacusutils}.

For larger $1024^3$ meshes with the same number of particles, the speedup is smaller but still substantial: 650 ms (15 M part/s) versus 2.5 sec (4 M part/s), both with 32 threads, for a factor of $3.8\times$.

As with the TSC computation, we test our results against \texttt{nbodykit} and find that they agree, up to minor differences due to different treatment of mode binning.


\subsection{Additional clustering plots}
\label{app:add}

In this appendix, we show the ZCV- and LCV-reduced clustering measurements for the two missing cases in the main body of the text: ELG power spectrum and LRG correlation function.

\begin{figure*}
    \centering
    \includegraphics[width=0.98\textwidth]{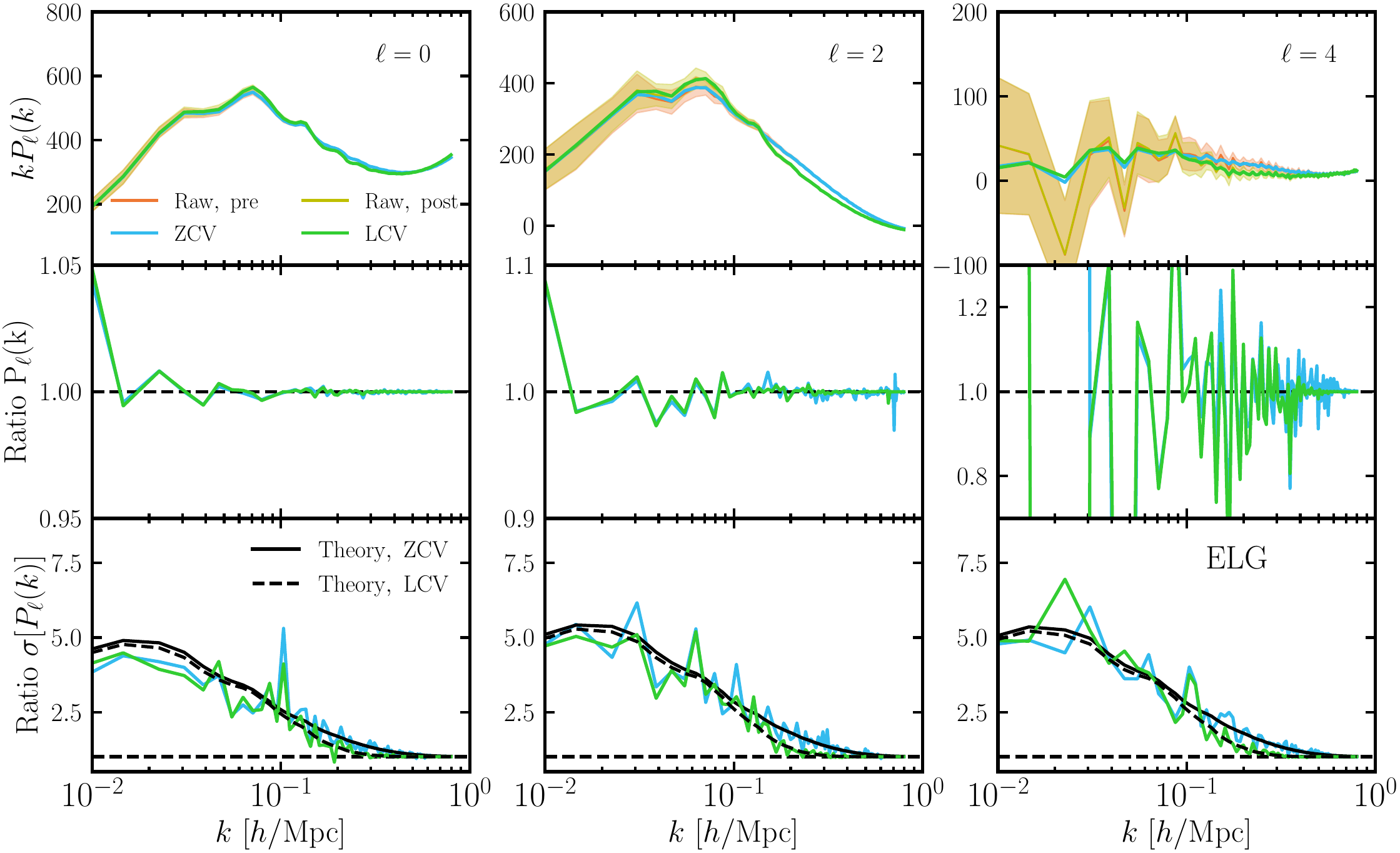} 
    \caption{Same as Fig.~\ref{fig:pk_lrg}, but for the ELG samples of Fig.~\ref{fig:xi_elg}. We see similar trends of reduction to the LRGs across all multipoles and correlated noise between LCV and ZCV due to the limitation of having only 25 cubic mocks. We find decent agreement with the theoretical prediction of the noise reduction with some slight noise noticeable in the monopole.}
    \label{fig:pk_elg}
\end{figure*}

\begin{figure*}
    \centering
    \includegraphics[width=0.98\textwidth]{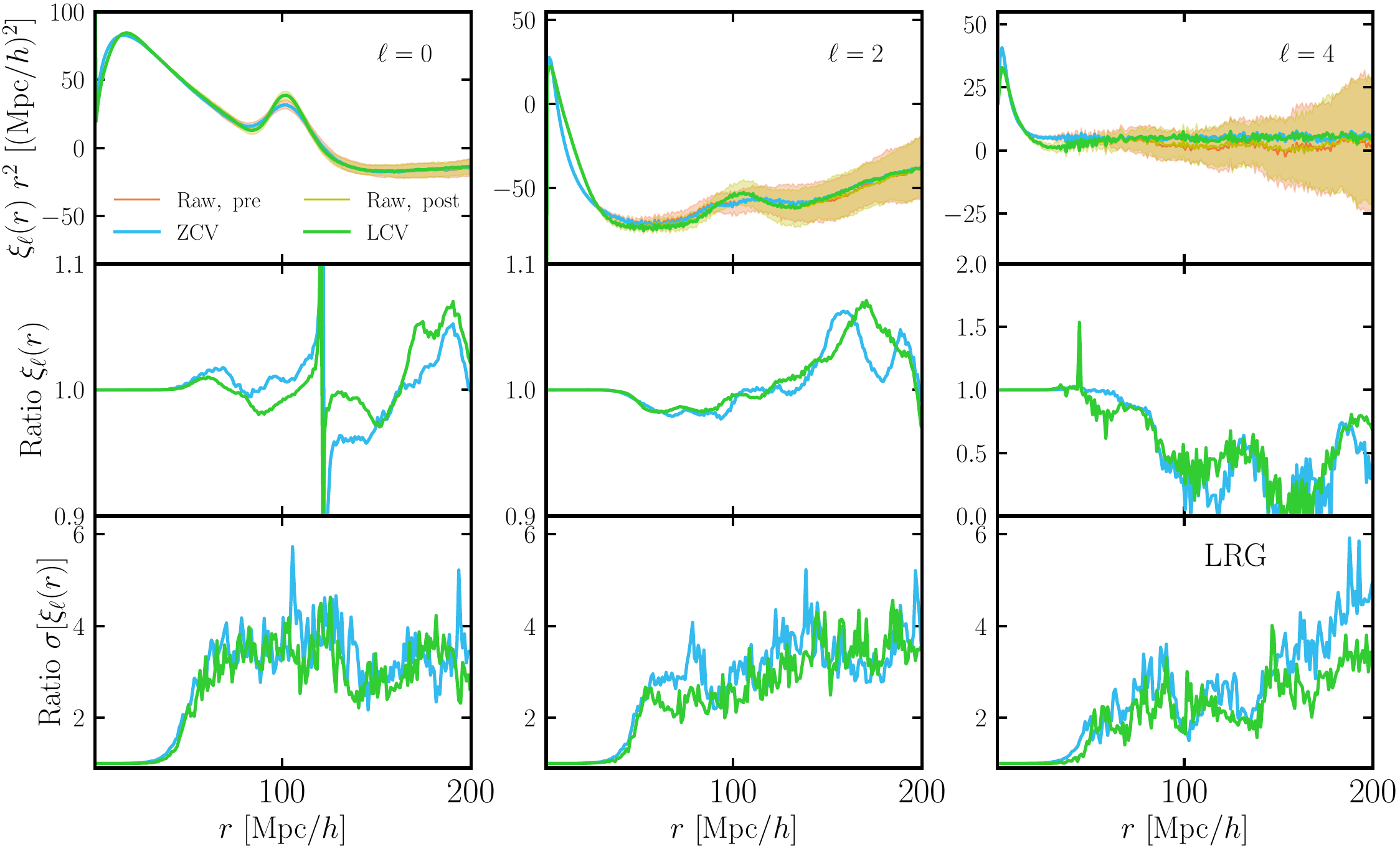} 
    \caption{Same as Fig.~\ref{fig:xi_elg}, but for the LRG samples of Fig.~\ref{fig:pk_lrg}. We see similar trends of reduction to the ELGs across all multipoles and correlated noise between LCV and ZCV due to the limitation of having only 25 cubic mocks.}
    \label{fig:xi_lrg}
\end{figure*}

\end{document}